\documentclass[global,twocolumn]{svjour}
\usepackage[T1]{fontenc}
\usepackage[latin9]{inputenc}
\usepackage{float}
\usepackage{amstext}
\usepackage{amssymb}
\usepackage{amsmath}
\usepackage{caption, ctable,threeparttable}
\usepackage{latexsym}
\usepackage{graphicx}
\begin{document}
\title{Random Unitary Evolution Model of Quantum Darwinism with pure decoherence}
\author{Nenad Balaneskovi$\textrm{\textrm{\ensuremath{\acute{\textrm{c}}}}}$%\inst{1} 
%\and Second author\inst{2}% etc
\thanks{\emph{email:} balaneskovic@gmx.net}%
}                     % Do not remove
\institute{Institut f\"ur Angewandte Physik, Technische Universit\"at Darmstadt, D-64289 Darmstadt, Germany %\and the second here
}
\date{Received: date / Revised version: date}
\maketitle
\begin{abstract}
We study the behavior of Quantum Darwinism (Zurek, \cite{key-3}) within the iterative, random unitary operations qubit-model of pure decoherence (Novotn$\textrm{\textrm{\ensuremath{\acute{\textrm{y}}}}}$ {\it{et al}}, \cite{key-1}). We conclude that Quantum Darwinism, which describes the quantum mechanical evolution of an open system $S$ from the point of view of its environment $E$, is not a generic phenomenon, but depends on the specific form of input states and on the type of $S$-$E$-interactions. Furthermore, we show that within the random unitary model the concept of Quantum Darwinism enables one to explicitly construct and specify artificial  input states of environment $E$ that allow to store information about an open system $S$ of interest with maximal efficiency. 
\end{abstract} %end of abstract
%
%---------------------------INTRODUCTION-------------------
\section{Introduction\label{A1}}

~~~From everyday-experience, classical states >>pre-exist<< objectively
and as such constitute >>classical reality<< in a sense that the
state of an open system $S$ can be measured and agreed upon by many
independent, mutually non-interacting observers, \emph{without being
disturbed}. This is done by intercepting fragments ($\equiv$ observers)
of the environment $E$ (indirect or non-demolition measurement \cite{key-0}).
Thus, one may ask: \emph{which sort of information about system $S$
is} redundantly\emph{ and }robustly \emph{memorized by numerous distinct
$E$-fragments, such that multiple observers may retrieve this same
information in a non-demolishing fashion, thereby confirming the effective
classicality of the $S$-state?}

Zurek's concept of Quantum Darwinism tries to answer the above question
by investigating what kind of information about system $S$ the environment
$E$ can store and proliferate in a stable, complete and redundant
way. It turns out that this redundantly stored information proliferated
throughout environment $E$ is the Shannon-entropy of the decohered
system $S$, that contains information about $S$-pointer states \cite{key-0_4,key-3}. 

These pointer-states, also known as interaction-robust $S$-states,
are those $S$-states most immune (invariant) towards numerous interactions
with the environment $E$. They are singled out by a characteristic
dynamical phenomenon, an interaction-induced decoherence, which explains
the process of destruction of quantum superpositions between states
of an open quantum system $S$ as a consequence of its interaction
with an environment $E$. Most decoherence-based explanations of the
emergence of classical $S$-states from quantum mechanical dynamics
deal solely with observations which can be made at the level of system
$S$, degrading its environment $E$ to the role of a >>sink<< that
carries away unimportant information about the preferred pointer-basis
of the observed system $S$ \cite{key-0}.

However, whereas the decoherence paradigm usually distinguishes between
an open system $S$ and its environment $E$, without specifying the
structure of the latter, Quantum Darwinism subdivides the environment
$E$ into \emph{non-overlapping subenvironments} (fragments or ``storage
cells'') accessible to measurements, that have already interacted
with system $S$ in the past and thus enclose Shannon-information
(entropy) about its preferred (pointer) states (i.e. $E$-registry
states are assumed to have a \emph{tensor product} structure). In
other words, Quantum Darwinism changes the perspective and regards
the environment $E$ as a large resource (>>quantum memory<<) which
could be used for indirect acquisition and storage of relevant information
about system $S$ and its pointer-basis (i.e. $E$ becomes a ``witness''
to the observed $S$-state) \cite{key-3}.

Accordingly, one can quantify the ``degree of objectivity'' of $S$-states
by simply counting the number of copies of their information record
in environment $E$. This number of copies of the information deposited
by a particular $S$-state into environmental fragments after many
$S$-$E$-interactions reveals its redundancy $R$. The higher the
$R$ of a particular $S$-state, the more ``classical'' it appears.

Similar to the Darwinistic concept ``survival of the fittest'',
the $S$-pointer states represent the ``fittest'' (``quasi-classical'')
states of an open system $S$ that survive numerous $S$-$E$-interactions
(measurements) long enough to deposit (imprint) multiple copies of
their information into environment $E$ \cite{key-0_2}. Ergo: high
information redundancy of $S$-states within environment $E$ implies
that information about the ``fittest'' observable (pointer state)
of system $S$ that survived constant monitoring by the environment
$E$ has been successfully distributed throughout all $E$-fragments,
enabling the environment to store redundant copies of information
about preferred system's observables and thus account for their objective
existence (``ein-selection'' \cite{key-0_1,key-0_3}).

In the following we intend to compare two qubit models of Quantum
Darwinism: Zurek's C(ontrolled-)NOT-evolution model \cite{key-3}
and the random unitary operations model \cite{key-1,key-2,key-4}
of an open $k$-qubit system $S$ interacting with an $n$-qubit environment
$E$. According to Zurek's qubit model the one qubit ($k=1$) open system
$S$ acts via CNOT-transformations as a control unit upon each of
the $n$ mutually non-interacting $E$-qubits (targets) \emph{only
once}. On the other hand, the random unitary evolution generalizes
Zurek's interaction procedure by iterating the directed graph (digraph)
of CNOT-interactions between a $k\geq1$ qubit system $S$ and mutually
non-interacting $E$-qubits, represented by the corresponding quantum
operation channel, $N\gg1$ times until the underlying dynamics forces
the input state $\hat{\rho}_{SE}^{in}$ of the entire system to converge
to the output state $\hat{\rho}_{SE}^{out}$. Such asymptotically
evolved $\hat{\rho}_{SE}^{out}$ can then be described by a subset
of the total Hilbert-space $\mathcal{H}_{SE}=\mathcal{H}_{S}\otimes\mathcal{H}_{E}$,
the so-called \emph{attractor space}, and attractor states therein. 

From the practical point of view, we want to answer two questions.
First: Which $\hat{\rho}_{SE}^{in}$ lead to Quantum Darwinism? Second:
Does Quantum Darwinism, and thus a perfect transfer of Shannon-entropy
into environment $E$, depend on a specific model being used, or is
it a model-independent phenomenon? Namely, since the random unitary
evolution can model systems $S$ subject to pure decoherence by singling
out the corresponding pointer states as a result of the asymptotic
iterative dynamics, it also enables one to specify (in comparison
with Zurek's model)\textbf{ }which types of input states $\hat{\rho}_{SE}^{in}$
store the ``classical'' Shannon information about system $S$ and
its pointer-basis efficiently into environment $E$. Finally, we also
want to use the random unitary model to see whether Quantum Darwinism
appears if we introduce into the corresponding interaction digraph
CNOT interactions between $E$-qubits.

This article is organized as follows: Section \ref{A2} deals with
basic physical and mathematical concepts of Quantum Darwinism (mutual
information, CNOT transformation, partial information plots, $S$-pointer
states) and discusses this phenomenon within the framework of Zurek's
qubit CNOT-evolution toy model \cite{key-3}. We thereby see that
for an open pure $k$-qubit $S$-input state $\hat{\rho}_{S}^{in}$
the CNOT transformation leads to Quantum Darwinism only if one starts
with $\hat{\rho}_{SE}^{in}=\hat{\rho}_{S}^{in}\otimes\hat{\rho}_{E}^{in}$
and $\hat{\rho}_{E}^{in}$ prepared as a pure $n$-qubit $E$-registry
state. In section \ref{A3} we first introduce the mathematical formalism
of iterated random unitary evolution \cite{key-1,key-2,key-4}. In
subsections \ref{A3.1} and \ref{A3.4} we show that introducing CNOT-interactions
between $E$-qubits suppresses the appearence of Quantum Darwinism.
In subsection \ref{A3.2} we present numerical results of the iterated
random unitary evolution, concluding that Zurek's qubit model of Quantum
Darwinism cannot be interpreted as a short-time limit ($\equiv$ small
number $N$ of iterations) of the random unitary evolution model. 

Then we turn our attention in subsection \ref{A3.3} to the asymptotic
($N\gg1$) behavior of mutual information within the random unitary
qubit model. This asymptotic behavior of mutual information of iterated,
random unitarily evolved output states $\hat{\rho}_{SE}^{out}$ allows
us to conclude that Quantum Darwinism and its appearence depends in
general on an underlying model used to describe interactions between
$S$- and $E$-qubits. Finally, we summarize the most important results
of our discussion before giving a brief outlook on interesting future
research problems connected with Quantum Darwinism (section \ref{A4}).
All detailed analytic calculations are given in 4 appendices: Appendix \ref{AAA}
displays output states $\hat{\rho}_{SE}^{out}$ of Zurek's CNOT-evolution
used in section \ref{A2}. Appendix \ref{AAB} explains why only the CNOT-transformation
leads to Quantum Darwinism, both in Zurek's and the random unitary
evolution model. In Appendix \ref{AAC} we derive (dimensionally) maximal and
minimal attractor subspaces that are used in the course of interpretation
of random unitarlly evolved $\hat{\rho}_{SE}^{out}$ in section \ref{A3}.
Finally, Appendix \ref{AAD_ges} contains a list of $\hat{\rho}_{SE}^{out}$
obtained by means of (dimensionally) maximal and minimal attractor
spaces that are necessary for the discussion of the random unitary
evolution in section \ref{A3}.
%---------------------------SECTION 2------------------------
\section{A qubit toy-model of Quantum Darwinism\label{A2}}

~~~In this section we briefly describe the simplest qubit model
of Quantum Darwinism, as suggested by Zurek \cite{key-3}, involving
an open pure $k=1$-qubit $S$ (given by the state vector $\left|\Psi_{S}^{in}\right\rangle =a\left|0\right\rangle +b\left|1\right\rangle $,
$\left(a,\, b\right)\in\mathbb{C}$ in the standard computational basis, where $|a|^{2}+|b|^{2}\stackrel{!}{=}1$
), which acts as a control-unit on its $\left(n\in\mathbb{N}\right)$-qubit
target (environment) $E\equiv\mathcal{E}_{1}\otimes\mathcal{E}_{2}\otimes...\otimes\mathcal{E}_{n}$.

Subsequently, we apply Zurek's qubit evolution model to different
input states $\hat{\rho}_{SE}^{in}$ of the total system and investigate
whether Quantum Darwinism appears within this model with respect to
different members of a one-parameter family of unitary transformations
that also encloses, as a special case, the unitary C(ontrolled)-NOT
operation.
According to Zurek's qubit model the interaction between system $S$
and environment $E$ has to occur as follows:
\begin{enumerate}
\item Start with a pure $k=1$-qubit open $\hat{\rho}_{S}^{in}=\left|\Psi_{S}^{in}\right\rangle \left\langle \Psi_{S}^{in}\right|$
and an arbitrary $n$-qubit $\hat{\rho}_{E}^{in}$, where $\hat{\rho}_{SE}^{in}=\hat{\rho}_{S}^{in}\otimes\hat{\rho}_{E}^{in}$.
\item Apply the CNOT-gate $\hat{U}_{CNOT}\left|i\right\rangle _{S}\left|j\right\rangle _{E}=\left|i\right\rangle _{S}\left|i\oplus j\right\rangle _{E}$
(where $\oplus$ denotes addition modulo $2$), such that the $S$-qubit
$i$ interacts successively and only once with each qubit $j$ of
$E$ until all $n$ $E$-qubits have interacted with $S$, resulting
in an entangled state $\hat{\rho}_{SE}^{out}$.
\item Trace out successively (for example from \emph{right to left}) $\left(n-L\right)$
qubits in $\hat{\rho}_{E}^{out}$ and $\hat{\rho}_{SE}^{out}$ - this
yields the $L$-qubit $\hat{\rho}_{E_{L}}^{out}$ and $\hat{\rho}_{SE_{L}}^{out}$,
with $0<L\leq n$, and an environmental fraction parameter $0<f=\frac{L}{n}\leq1$.
\item Compute the eigenvalue spectra $\left\{ \lambda_{1},...,\lambda_{d\left(f\right)}\right\} $
of $\hat{\rho}_{S}^{out}$, $\hat{\rho}_{E_{f}}^{out}$ and $\hat{\rho}_{SE_{f}}^{out}$
and the $f$-dependent von Neumann entropies \[H\left(\hat{\rho}\left(f\right)\right)=-\underset{i=1}{\overset{d\left(f\right)}{\sum}}\lambda_{i}\log_{2}\lambda_{i}\geq0,\,\underset{i=1}{\overset{d\left(f\right)}{\sum}}\lambda_{i}\overset{!}{=}1\]
(where $d\left(f\right)$ is the dimensionality of $\hat{\rho}\left(f\right)$ in question). 
\item Divide all entropies by $H\left(S_{class}\right)$ to obtain the ratio
$I\left(S:\, E_{f}\right)/H\left(S_{class}\right)$ depending on the $E$-fraction parameter $f$, with \textbf{mutual information} (MI)
\begin{equation}
I\left(S:\, E_{f}\right)=H\left(S\right)+H\left(E_{f}\right)-H\left(S,\, E_{f}\right),\label{Gl 3.3}
\end{equation}
 that quantifies the amount of the proliferated Shannon entropy (>>classical
information<<) \cite{key-0_2,key-3_1} 
\begin{equation}
H\left(S_{class}\right)=-\sum_{i}p_{i}\log_{2}p_{i}=H\left(\left\{ \left|\pi_{i}\right\rangle \right\} \right),\label{Gl 3.2}
\end{equation}
where probabilities $p_{i}=\mathbf{Tr}_{E}\left\langle \pi_{i}\right|\hat{\rho}_{SE}^{class}\left|\pi_{i}\right\rangle $
emerge as partial traces of an effectively decohered (>>quasi classical<<)
$S$-state $\hat{\rho}_{S}^{class}$ w.r.t. the particular $S$-pointer-basis
$\left\{ \left|\pi_{i}\right\rangle \right\} $, and the redundancy
\begin{equation}
\begin{array}{c}
R=1/f^{*}\,\left(0<f^{*}\leq1\right)\\
\textrm{with}\, I\left(S:\, E_{f=f^{*}}\right)\approx H\left(S_{class}\right)\left(n\gg1\right)
\end{array}\label{Gl 3.4}
\end{equation}
of the measured $\left\{ \left|\pi_{i}\right\rangle \right\} $ in
the limit $n\gg1$ of effective decoherence. 
\item Finally, plot $I\left(S:\, E_{f}\right)/H\left(S_{class}\right)$
vs $0<f\leq1$ \\(\textbf{Partial Information Plot} (\textbf{PIP}) of
MI).
\end{enumerate}

Now we look at the specific input state \[\hat{\rho}_{SE}^{in}=\left|\Psi_{S}^{in}\right\rangle \left\langle \Psi_{S}^{in}\right|\otimes\left|0_{n}\right\rangle \left\langle 0_{n}\right|\]
with $\left|0_{n}\right\rangle \equiv\left|0\right\rangle ^{\otimes n}$
(ground state $\hat{\rho}_{E}^{in}$) \cite{key-3}. Let the one $S$-qubit
transform each $E$-qubit via CNOT \emph{only once} until the entire
environment $E$ is affected, giving $\forall L>0$
\begin{equation}
\left|\Psi_{SE_{L=n}}^{out}\right\rangle =a\left|0\right\rangle \otimes\left|0_{L=n}\right\rangle +b\left|1\right\rangle \otimes\left|1_{L=n}\right\rangle,\label{Gl 3.5}
\end{equation}
with von Neumann-entropies
\begin{align*}
\begin{split}
H\left(S,\, E_{L}\right)&=H\left(S_{class}\right)\cdot\left(1-\delta_{L,n}\right)\\
H\left(E_{L}\right)&=H\left(S\right)=H\left(S_{class}\right)\\
H\left(S_{class}\right)&=-\left|a\right|^{2}\log_{2}\left|a\right|^{2}-\left|b\right|^{2}\log_{2}\left|b\right|^{2}.
\end{split}
\end{align*}
(\ref{Gl 3.5}) shows that $I\left(S:\, E_{f}\right)$, after the
$L$-th $E$-qubit has been taken into account, increases from zero
to the value \[I\left(S:\, E_{f}\right)\equiv H\left(S_{class}\right)\Rightarrow I\left(S:\, E_{f}\right)/H\left(S_{class}\right)=1,\]
implying that each fragment (qubit) of environment $E$ supplies complete
information about the $S$-pointer observables $\left\{ \left|\pi_{i}\right\rangle \right\} $.
Since the very first CNOT-operation forces the system $S$ to decohere
completely into its pointer basis $\left\{ \left|\pi_{i}\right\rangle \right\} \equiv\left\{ \left|0\right\rangle ,\,\left|1\right\rangle \right\} $,
one encounters the influence of Quantum Darwinism on system $S$:
from all possible $S$-states, which started its dynamics within a
pure $\hat{\rho}_{S}^{in}$, only diagonal elements survive constant
monitoring of environment $E$, whereas off-diagonal elements of $\hat{\rho}_{S}^{in}$
vanish due to decoherence, i.e. monitoring of system $S$ by its environment
$E$ selects a preferred $\left\{ \left|\pi_{i}\right\rangle \right\} $,
leading to a continued increase of its $R$ throughout the environment
$E$.
 
\begin{figure}[H]
\resizebox{0.48\textwidth}{!}{%natwidth=610,natheight=642
  \includegraphics{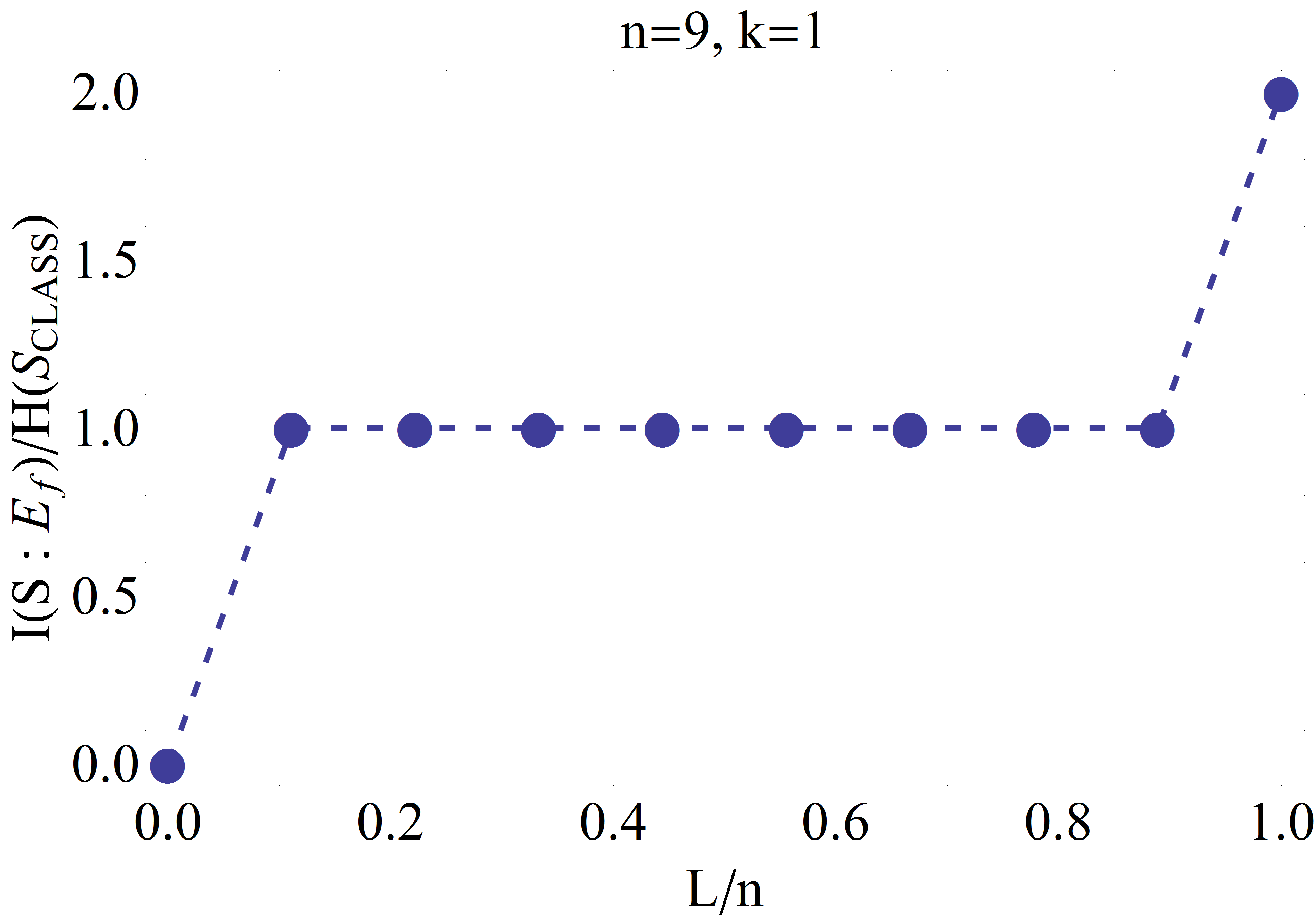}
}
\caption{PIP of MI and $R$ of $H\left(\hat{\rho}_{S}^{out}\right)= H\left(S_{class}\right)$
stored in $E$ w.r.t. $0<f=L/n\leq1$, $k=1$ qubit pure $\hat{\rho}_{S}^{in}$,
$\hat{\rho}_{E}^{in}=\left|0_{n}\right\rangle \left\langle 0_{n}\right|$,
$\hat{\rho}_{SE}^{in}=\hat{\rho}_{S}^{in}\otimes\hat{\rho}_{E}^{in}$
in (\ref{Gl 3.5}), after $\widehat{U}_{\textrm{CNOT}}$-evolution
in accord with Zurek's model \cite{key-3}. \label{Fig.1}}
\end{figure}

After decoherence we obtain \[I\left(S:\, E_{f}\right)=H\left(S_{class}\right)=H\left(E_{f}\right)=H\left(S,\, E_{f}\right),\]
valid for any $E$-fragment, as long $f<1$. After inclusion of the
entire environment $E$ ($f=1$) we obtain the maximum \[I\left(S:\, E\right)=2H\left(S_{class}\right)\]
of MI (>>quantum peak<<, accessible through global measurements
of $\hat{\rho}_{S,E}^{out}$ due to $H\left(S,\, E_{f=1}\right)=0$).
Since each $E$-qubit in (\ref{Gl 3.5}) is assumed to contain a perfect
information replica about $\left\{ \left|\pi_{i}\right\rangle \right\} $,
its $R$ is given by the number of qubits in the environment $E$,
e.g. $R=n$. 
This constrains the form of MI in its PIP (see Fig. \ref{Fig.1}),
which jumps from $0$ to $H\left(S_{class}\right)$ of $S$ at $f=f^{*}=1/n$,
continues along the 'plateau' until $f=1-1/n$, before it eventually
jumps up again to $2H\left(S_{class}\right)$ at $f=1$.

$I\left(S:\, E_{f}\right)/H\left(S_{class}\right)\geq1$
indicates high $R$ (objectivity) of $H\left(S_{class}\right)$ proliferated
throughout $E$. Also, by intercepting already one $E$-qubit we can
reconstruct $\left\{ \left|\pi_{i}\right\rangle \right\} $, regardless
of the order in which the $n$ $E$-qubits are being successively
traced out. Only if we need a small fraction of the environment $E$
enclosing maximally $n\cdot f^{*}=k\ll n$ $E$-qubits \cite{key-3},
to reconstruct $\left\{ \left|\pi_{i}\right\rangle \right\} $, Quantum
Darwinism appears: i.e., it is not only important that the PIP-'plateau'
appears, more relevant is its length $1/f^{*}\equiv R$ of $\left\{ \left|\pi_{i}\right\rangle \right\} $.
%0.48\textwidth
The main question we aim to address w.r.t. Zurek's and the random
unitary operations model is:
Which types of input states $\hat{\rho}_{SE}^{in}$ validate the relation
\begin{equation}
\frac{I\left(S:\, E_{f}\right)}{H\left(S_{class}\right)}=\frac{H\left(S\right)+H\left(E_{f}\right)-H\left(S,\, E_{f}\right)}{H\left(S_{class}\right)}\geq1,\label{Gl 3.5.1}
\end{equation}
with $H\left(S\right)\approx H\left(S_{class}\right)$ and $H\left(E_{f}\right)\geq H\left(S,E_{f}\right)$
at least for all $\left(k\leq L\leq [n\gg1]\right)$, regardless of
the order in which the $n$ $E$-qubits are being successively traced
out from the output state $\hat{\rho}_{SE}^{out}$?
In order to answer this question we discuss in the following the $f$-dependence of MI for different $\hat{\rho}_{SE}^{in}$ from the point of view
of Zurek's qubit model. Fig. \ref{Fig.2} below displays the behavior
of the MI vs $f$ for different $\hat{\rho}_{E}^{in}$ from Tab. \ref{Tab. A1} in Appendix \ref{AAA} which justifies the following conclusions:

It is in general important in which order one traces out $E$-qubits
from the output-state $\hat{\rho}_{SE_{L}}^{out}$, as indicated by
the $\bullet$-dotted curve (Quantum Darwinism appears) and the $\blacksquare$-dotted
curve (no Quantum Darwinism, since the relation $H\left(E_{L}\right)=H\left(S,\, E_{L}\right)$
holds only $\forall\left(k<L\leq n\right)$, but not for $L=k$) in Fig. \ref{Fig.2}:
when tracing out $E$-qubits as in (\ref{Gl 3.5}) from right to left
Quantum Darwinism appears only if, for each fixed value of $\left(k\leq L<n\right)$,
$\hat{\rho}_{SE_{L}}^{out}$ acquires the structure displayed in (\ref{Gl 3.5}),
that emerges when starting the CNOT-evolution with a pure, $n$-qubit
registry input state of the environment $\hat{\rho}_{E}^{in}=\left|0_{n}\right\rangle \left\langle 0_{n}\right|$.

Introducing classical correlations into $\hat{\rho}_{SE}^{in}=\hat{\rho}_{S}^{in}\otimes\hat{\rho}_{E}^{in}$
(with a one-qubit pure $\hat{\rho}_{S}^{in}$) by writing $\hat{\rho}_{E}^{in}$
as a convex sum of pure $n$-qubit registry states $ \left|y\right\rangle \left\langle y\right|$, with $y\in\left\{ 0,\,...,\,2^{n}-1\right\}$ (rank one operators)
in the standard computational basis, tends in general to suppress the
appearance of the MI-plateau: in case of a totally mixed $\hat{\rho}_{E}^{in}$
the MI is even zero $\forall\left(0<L<n\right)$, as indicated by
the $\blacklozenge$-dotted and $\blacktriangle$-dotted curves in
Fig. \ref{Fig.2}. Quantum correlations within $\hat{\rho}_{E}^{in}$do
not improve the situation, but lead in general to the relation $H\left(S\right)=H\left(\hat{\rho}_{S}^{out}\right)<H\left(S_{class}\right)$
instead, as shown by the $\blacktriangledown$-dotted curve in Fig.
\ref{Fig.2}.

We can extend Zurek's interaction algorithm to systems $S$ with more
than one qubit ($k>1$) by assuming the environment $E$ to contain $n'=k\cdot n\gg k$
qubit-cells (i.e. one subdivides $n'$ $E$-qubits into $k$ disjoint
subsets %$E\equiv\underset{i=1}{\overset{k}{\bigcup}}E_{i}$ 
$E\equiv\bigcup_{i=1}^{k} E_{i}$ with $\left|E_{i}\right|=n\,\forall i$)
and allowing each $S$-qubit to interact with only one $E_{i}$-subset
of environment $E$ and only once with each of the $n\gg k$ $E$-qubits
within the particular $E_{i}$. 

Then, with $\hat{\rho}_{SE}^{in}=\hat{\rho}_{S}^{in}\otimes\hat{\rho}_{E}^{in}$, for
$\hat{\rho}_{E}^{in}=\left|0_{n'}\right\rangle \left\langle 0_{n'}\right|$
and a pure two-qubit ($k=2$) state $\hat{\rho}_{S}^{in}$, the corresponding
PIP is given for all $\left(k=2\leq L\leq n'-k\right)$ by the $\bullet$-dotted
plateau in Fig. \ref{Fig.2}, whereas\footnote{the lower bound follows from the trivial initial probability distribution$\left\{ \left|a_{i}\right|^{2}=1,\,\left|a_{i'}\right|^{2}=0\,\forall i'\neq i\right\} $,
whereas the upper bound emerges from $\left\{ \left|a_{i}\right|^{2}=2^{-k}\right\} _{i=0}^{2^{k}-1}\forall i$
in $\hat{\rho}_{S}^{in}$.%
} \[
\begin{array}{c}
\ensuremath{H\left(S,\, E_{L}\right)<H\left(E_{L}\right)},\,\,\quad\ensuremath{n'-k+1\leq L\leq n'}\\
\ensuremath{0\leq I\left(S:\, E_{L=1}\right)/H\left(S_{class}\right)\leq0.5},\,\,\quad\ensuremath{0<L<k=2}.
\end{array}
\]

Thus, in Zurek's pure decoherence qubit-model of Quantum Darwinism
the specified CNOT-evolution yields the MI-plateau also for pure $\hat{\rho}_{S}^{in}$
with $k>1$ qubits if we start its evolution within $\hat{\rho}_{SE}^{in}=\hat{\rho}_{S}^{in}\otimes\hat{\rho}_{E}^{in}$
and with a pure one registry $n'$-qubit state $\hat{\rho}_{E}^{in}=\left|y\right\rangle \left\langle y\right|$
in the standard computational basis ($y\in\left\{ 0,\,...,\,2^{n'}-1\right\}$).

Certainly, if we deliberately design $\hat{\rho}_{SE}^{in}$ such
that it remains unaltered under the CNOT-evolution by entangling the
pointer-basis $\left\{ \left|\pi_{i}\right\rangle \right\} \equiv\left\{ \left|0\right\rangle ,\,\left|1\right\rangle \right\} $
of a $k=1$ qubit system $S$ with one-qubit $E$-eigenstates $\left|s_{1}\right\rangle =2^{-1/2}\left(\left|0\right\rangle +\left|1\right\rangle \right)$ and $\left|s_{2}\right\rangle =2^{-1/2}\left(\left|0\right\rangle -\left|1\right\rangle \right)$
of the CNOT-transformation (Pauli matrix) $\hat{\sigma}_{x}$ according
to
\begin{equation}
\begin{array}{c}
\left|\Psi_{SE}^{in}\left(L=n\right)\right\rangle =a\left|0\right\rangle \otimes\left|s_{1}^{L=n}\right\rangle +b\left|1\right\rangle \otimes\left|s_{2}^{L=n}\right\rangle \\
\left|s_{m}^{L}\right\rangle =\left|s_{m}\right\rangle ^{\otimes L},\,\hat{\sigma}_{x}\left|s_{m}\right\rangle =\underset{:=\lambda}{\underbrace{\left(-1\right)^{m+1}}}\left|s_{m}\right\rangle 
\end{array}\label{Gl 3.6}
\end{equation}
(with $m\in\left\{ 1,\,2\right\} $, $\left\langle s_{1}|s_{2}\right\rangle =0$),
(\ref{Gl 3.6}) would lead to the PIP displayed in Fig. \ref{Fig.1}:
i.e. Quantum Darwinism would appear. One can even show that (\ref{Gl 3.6})
leads to Quantum Darwinism only for $k=1$ qubit system $S$ (s. Appendix \ref{AAB}).
\begin{figure}[H]
\includegraphics[width=0.48\textwidth]{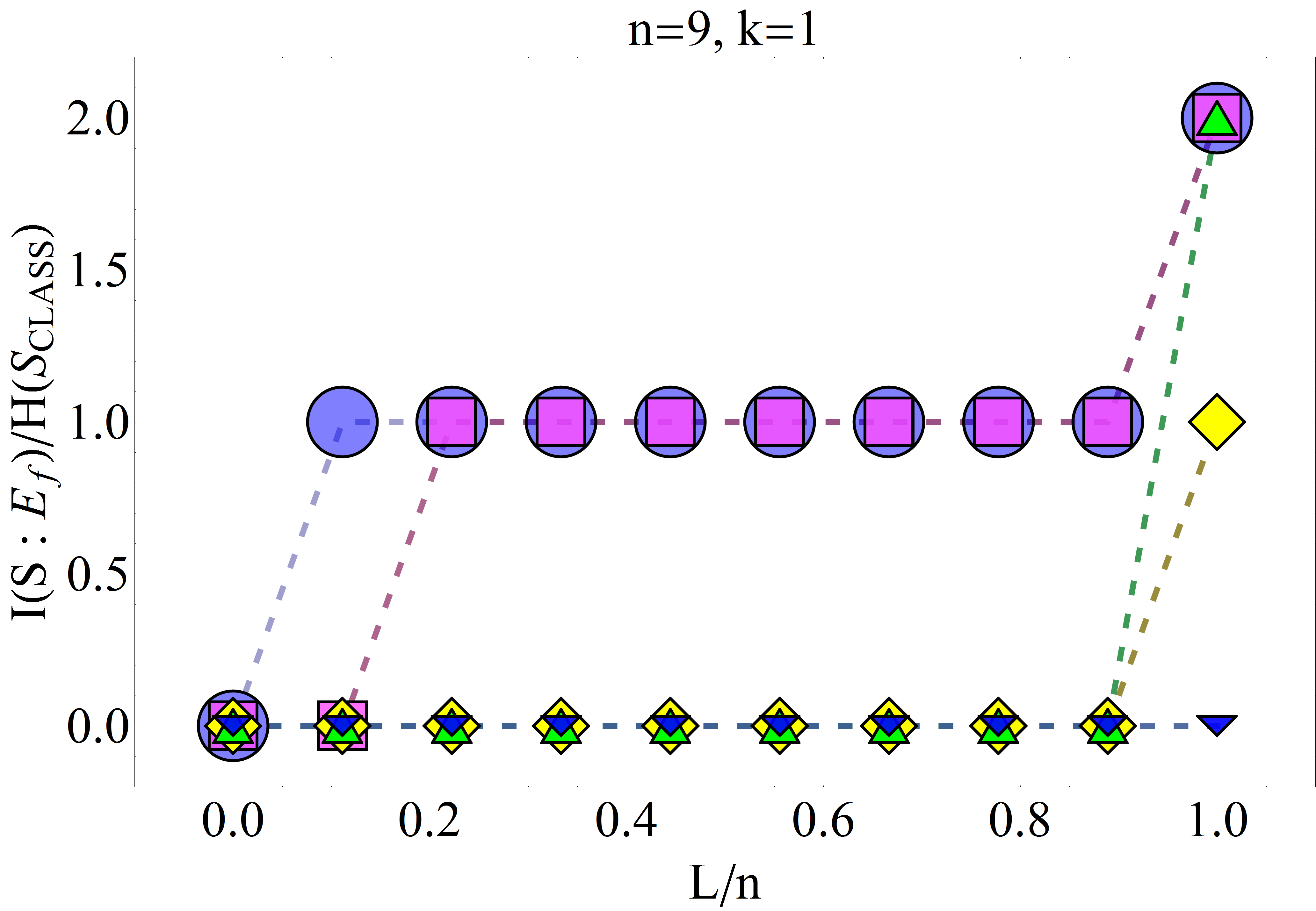}
\caption{PIP for $\hat{\rho}_{SE}^{out}$ from $\hat{\rho}_{SE}^{in}=\left|\Psi_{S}^{in}\right\rangle \left\langle \Psi_{S}^{in}\right|\otimes\hat{\rho}_{E}^{in}$
(where $k=1$, $n=9$, $\left|\Psi_{S}^{in}\right\rangle =a\left|0\right\rangle +b\left|1\right\rangle $)
and different $\hat{\rho}_{E}^{in}$ (s. also Tab. \ref{Tab. A1}
in Appendix \ref{AAA}) in Zurek's qubit model.\label{Fig.2}}
\end{figure}
%
%-------------------------SECTION 3------------------
\section{Random Unitary Model of Quantum Darwinism\label{A3}}

~~~In the present section we summarize the iterative evolution
formalism of the random unitary model before discussing its most important
results regarding Quantum Darwinism in subsections \ref{A3.1}-\ref{A3.4}. 

\begin{figure}[H]
\center\includegraphics[width=0.20\textwidth]{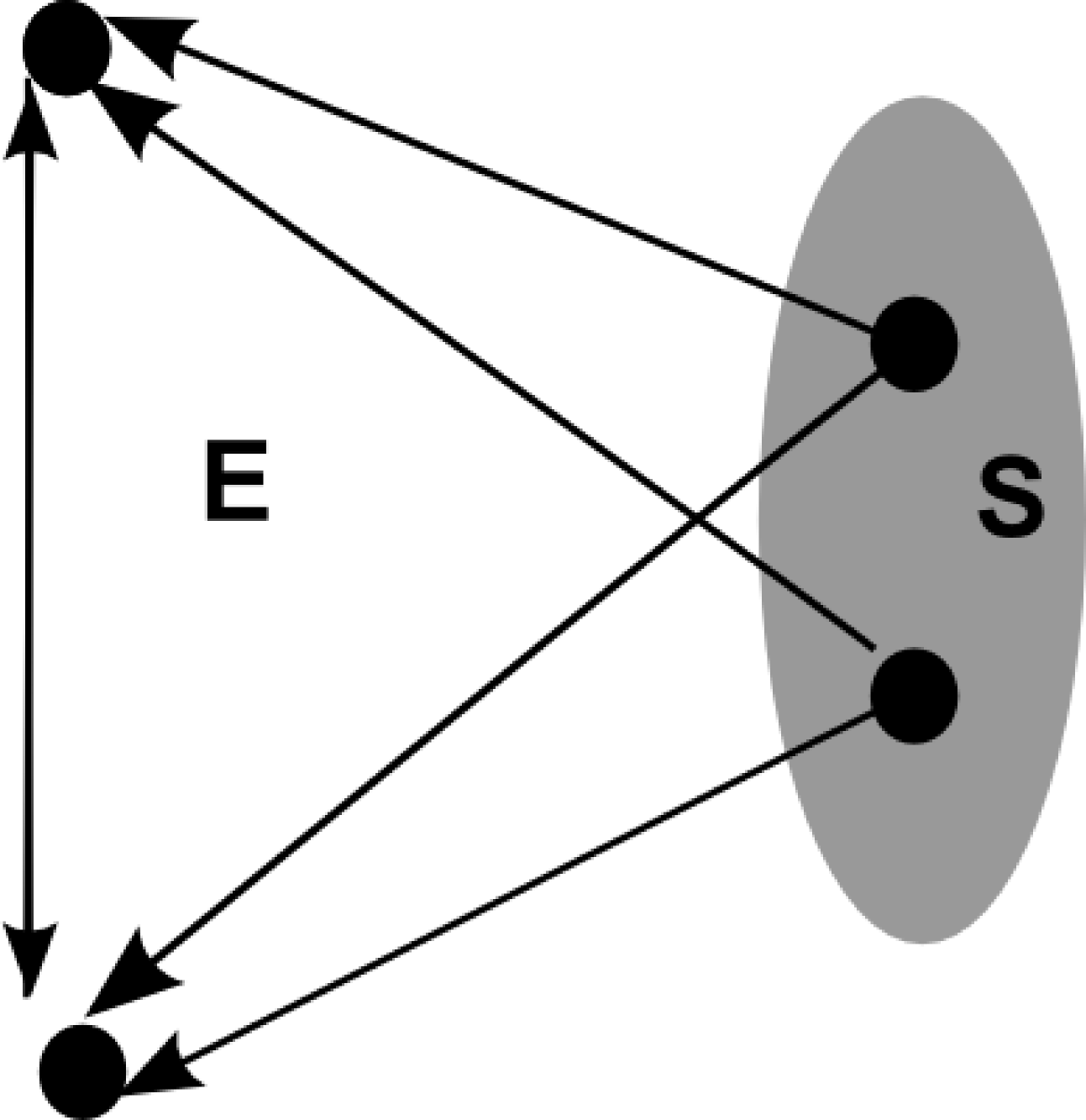}

\caption{Interaction digraph (ID) between system $S$ and environment $E$ with pure decoherence
within the random unitary model \cite{key-2}. \label{Fig.3}}
\end{figure}

Random unitary operations can model the pure decoherence of open system
$S$ with $k$ qubits (control, index $i$) interacting with $n$
$E$-qubits (targets $j$) (as indicated in the directed interaction
graph (digraph) in Fig. \ref{Fig.3}) by the one-parameter family
of two-qubit \textquoteright{}controlled-U\textquoteright{} unitary
transformations (in the standard one-qubit computational basis $\left\{ \left|0\right\rangle ,\,\left|1\right\rangle \right\} $) 

\begin{equation}
\widehat{U}_{ij}^{\left(\phi\right)}=\left|0\right\rangle _{i}\left\langle 0\right|\otimes\widehat{I}_{1}^{\left(j\right)}+\left|1\right\rangle _{i}\left\langle 1\right|\otimes\widehat{u}_{j}^{\left(\phi\right)}\label{Gl 2.1}
\end{equation}
(where $\widehat{I}_{1}^{\left(j\right)}=\left|0\right\rangle _{j}\left\langle 0\right|+\left|1\right\rangle _{j}\left\langle 1\right|$).
(\ref{Gl 2.1}) indicates that only if an $S$-qubit $i$ should be
in an excited state, the corresponding targeted $E$-qubit $j$ hast
to be modified by a $\left(0\leq\phi\leq\pi\right)$-parameter $\widehat{u}_{j}^{\left(\phi\right)}$
\cite{key-2} (with Pauli matrices $\hat{\sigma}_{l}$, $l\in\left\{ x,\, y,\, z\right\} $)
\begin{equation}
\widehat{u}_{j}^{\left(\phi\right)}=\hat{\sigma}_{z}^{\left(j\right)}\cos\phi+\hat{\sigma}_{x}^{\left(j\right)}\sin\phi\Rightarrow\widehat{u}_{j}^{\left(\phi=\pi/2\right)}=\hat{\sigma}_{x}^{\left(j\right)},\label{Gl 2.2}
\end{equation}
which for $\phi=\pi/2$ yields the CNOT-gate \cite{key-1,key-2,key-4}.
Arrows of the interaction digraph (ID) in Fig. \ref{Fig.3} from $S$-
to $E$-qubits represent two-qubit interactions $\widehat{u}_{j}^{\left(\phi\right)}$
between randomly chosen qubits $i$ and $j$ with probability distribution
$p_{e}$ used to weight the edges $e=(ij)\in M$ of the digraph ($E$-qubits
are in general allowed to interact among themselves).

All interactions are well separated in time. The $S$-qubits do not
interact among themselves. In order to model the decoherence-induced
measurement process of system $S$ by environment $E$ we let an input
state $\hat{\rho}_{SE}^{in}$ evolve by virtue of the following iteratively
applied random unitary quantum operation (completely positive unital map)
$\mathcal{P}\left(\centerdot\right)\equiv\sum_{e\in M}K_{e}\left(\centerdot\right)K_{e}^{\dagger}$
(with Kraus-operators given by $K_{e}:=\sqrt{p_{e}}\widehat{U}_{e}^{\left(\phi\right)}$)
\cite{key-1,key-2,key-4}:

1. The quantum state $\hat{\rho}(N)$ after $N$ iterations is changed
by the $(N+1)$-th iteration to the quantum state (quantum Markov
chain)
\begin{equation}
\widehat{\rho}\left(N+1\right)=\sum_{e\in M}p_{e}\widehat{U}_{e}^{\left(\phi\right)}\widehat{\rho}\left(N\right)\widehat{U}_{e}^{\left(\phi\right)\dagger}\equiv\mathcal{P}\left(\widehat{\rho}\left(N\right)\right).\label{Gl 2.5}
\end{equation}

2. In the asymptotic limit $N\gg1$ $\widehat{\rho}\left(N\right)$
is independent of $\left(p_{e},\, e\in M\right)$ and determined by
linear attractor spaces $\mathcal{A}_{\lambda}\subset\mathcal{H}_{SE}$,
as subspaces of the total $S$-$E$-Hilbert space $\mathcal{H}_{SE}=\mathcal{H}_{S}\otimes\mathcal{H}_{E}$
to the eigenvalues $\lambda$ (with $\left|\lambda\right|=1$), that
contain mutually orthonormal solutions (states) $\hat{X}_{\lambda,i}$
of the eigenvalue equation \cite{key-1,key-4}
\begin{equation}
\widehat{U}_{e}^{\left(\phi\right)}\widehat{X}_{\lambda,i}\widehat{U}_{e}^{\left(\phi\right)\dagger}=\lambda\widehat{X}_{\lambda,i},\,\forall\, e\in M.\label{Gl 2.6}
\end{equation}

3. For known attractor spaces $\mathcal{A}_{\lambda}$ we get from
an initial state $\hat{\rho}_{SE}^{in}$ the resulting $S$-$E$-state
$\widehat{\rho}_{SE}^{out}=\widehat{\rho}_{SE}\left(N\gg1\right)$
spanned by $\widehat{X}_{\lambda,i}$
\begin{equation}
\widehat{\rho}_{SE}^{out}=\mathcal{P}^{N}\left(\widehat{\rho}_{SE}^{in}\right)=\underset{\left|\lambda\right|=1,\, i=1}{\overset{d^{\lambda}}{\sum}}\lambda^{N}\textrm{Tr}\left\{ \widehat{\rho}_{SE}^{in}\widehat{X}_{\lambda,i}^{\dagger}\right\} \widehat{X}_{\lambda,i},\label{Gl 2.7}
\end{equation}
where $d^{\lambda}$ denotes the dimensionality of the attractor space
$\mathcal{A}_{\lambda}$ w.r.t. the eigenvalue $\lambda$.
\subsection{Minimal attractor space\label{A3.1}}

~~~What happens with Quantum Darwinism in the framework of the
random unitary evolution model if we let the $E$-qubits interact
with each other? From \cite{key-1} we know that an ID with all mutually
interacting $E$-qubits leads to the minimal attractor space structure
(\ref{Gl G8}) of Appendix \ref{AAC} associated with an eigenvalue $\lambda=1$ of (\ref{Gl 2.7}).
However, for this minimal $\lambda=1$ attractor subspace of (\ref{Gl 2.7})
to emerge one does not need to insist that \emph{all} $n$ $E$-qubits
should interact with each other. It suffices to have a strongly connected
ID that contains a closed arrow path between $n$ $E$-qubits \cite{key-1,key-2,key-4}.
However, $2n-3$ is the critical (and
maximal) number of $\widehat{u}_{j}^{\left(\phi\right)}$-bindings
that one may insert between $E$-qubits into the ID of Fig. \ref{Fig.3}
and still avoid the minimal $d^{\lambda=1}$ from (\ref{Gl G6}) of Appendix \ref{AAC}.
Thus, for $\geq2\left(n-1\right)$ $\widehat{u}_{j}^{\left(\phi\right)}$-bindings
between $E$-qubits the corresponding ID remains strongly connected,
leading always to the minimal $\lambda=1$ attractor space (\ref{Gl G8}) of Appendix \ref{AAC},
whereas the $\lambda=-1$ attractor space of (\ref{Gl 2.7}) vanishes
already after inserting a single interaction arrow into environment
$E$ (s. Appendix \ref{AAC}). Here we first turn to the physical interpretation
of (\ref{Gl G8}) from Appendix \ref{AAC}.

\subsubsection{State structure of the attractor space\label{A3.1.1}}

~

The main differences between the maximal and the minimal $\lambda=1$
attractor subspace (s. (\ref{Gl G4}) and (\ref{Gl G8}) in Appendix \ref{AAC})
that mainly determine the process of decoherence and transfer of $H\left(S_{class}\right)$
to $E$ are two-fold: 1) within the minimal $\lambda=1$ attractor
subspace (\ref{Gl G8}) only the ground $E$-registry state $\left|0_{n}\right\rangle $
appears, whereas in (\ref{Gl G4}) of the maximal $\lambda=1$ attractor
subspace all $2^{n}$ $E$-registry states contribute; 2) On the other
hand, in (\ref{Gl G4}) the $E$-registry states $|y\rangle$ are
correlated within the diagonal $S$-subspace $\left|0_{k}\right\rangle \left\langle 0_{k}\right|$
only with each other, whereas (\ref{Gl G8}) also allows the remaining
$E$-registry state $|0_{n}\rangle$ to be correlated with the $\widehat{u}_{j}^{\left(\phi\right)}$-symmetry
state $\left|s_{c_{1}}^{n}\right\rangle $. This means that effectively
the contribution of the $S$-subspace $\left|0_{k}\right\rangle \left\langle 0_{k}\right|$
in (\ref{Gl G8}) to $I\left(S:\, E_{L}\right)$, contrary to (\ref{Gl G4}),
becomes exponentially suppressed due to $\left|s_{c_{1}}^{n}\right\rangle $.
The implications of this exponential, decoherence induced suppression
of $S$-subspace $\left|0_{k}\right\rangle \left\langle 0_{k}\right|$
in (\ref{Gl G8}) regarding Quantum Darwinism will be discussed in
the forthcoming subsection.

\subsubsection{Results of the CNOT-evolution\label{A3.1.2}}

~

Decomposing $\hat{\rho}_{SE}^{in}$ for $n\gg k$ $S$-qubits by means
of (\ref{Gl 2.7}) and (only) linear independent $\widehat{X}_{\lambda,i}$
from (\ref{Gl G8}) of Appendix \ref{AAC}, after first orthonormalizing all $\widehat{X}_{\lambda,i}$
in accord with the Gram-Schmidt algorithm, we obtain the CNOT-asymptotically
evolved $\hat{\rho}_{SE}^{out}$ displayed in (\ref{Gl J1})-(\ref{Gl J5})
of Appendix \ref{AAD2-1}. We consider in the following different inputs $\hat{\rho}_{SE}^{in}$
of the random unitary evolution and their PIPs obtained from the corresponding
outputs $\hat{\rho}_{SE}^{out}$.

\paragraph{I) $\hat{\rho}_{SE}^{in}=|\Psi_{S,\,k}^{in}\rangle\langle
\Psi_{S,\,k}^{in}|\otimes\hat{\rho}_{E}^{in}$,
$\hat{\rho}_{E}^{in}=\left|y\right\rangle \left\langle y\right|$, $\ensuremath{y\in\left\{ 0_{n},\,1_{n}\right\} }$ 
}

~

\begin{figure}[H]
\includegraphics[width=0.48\textwidth,natwidth=610,natheight=642]{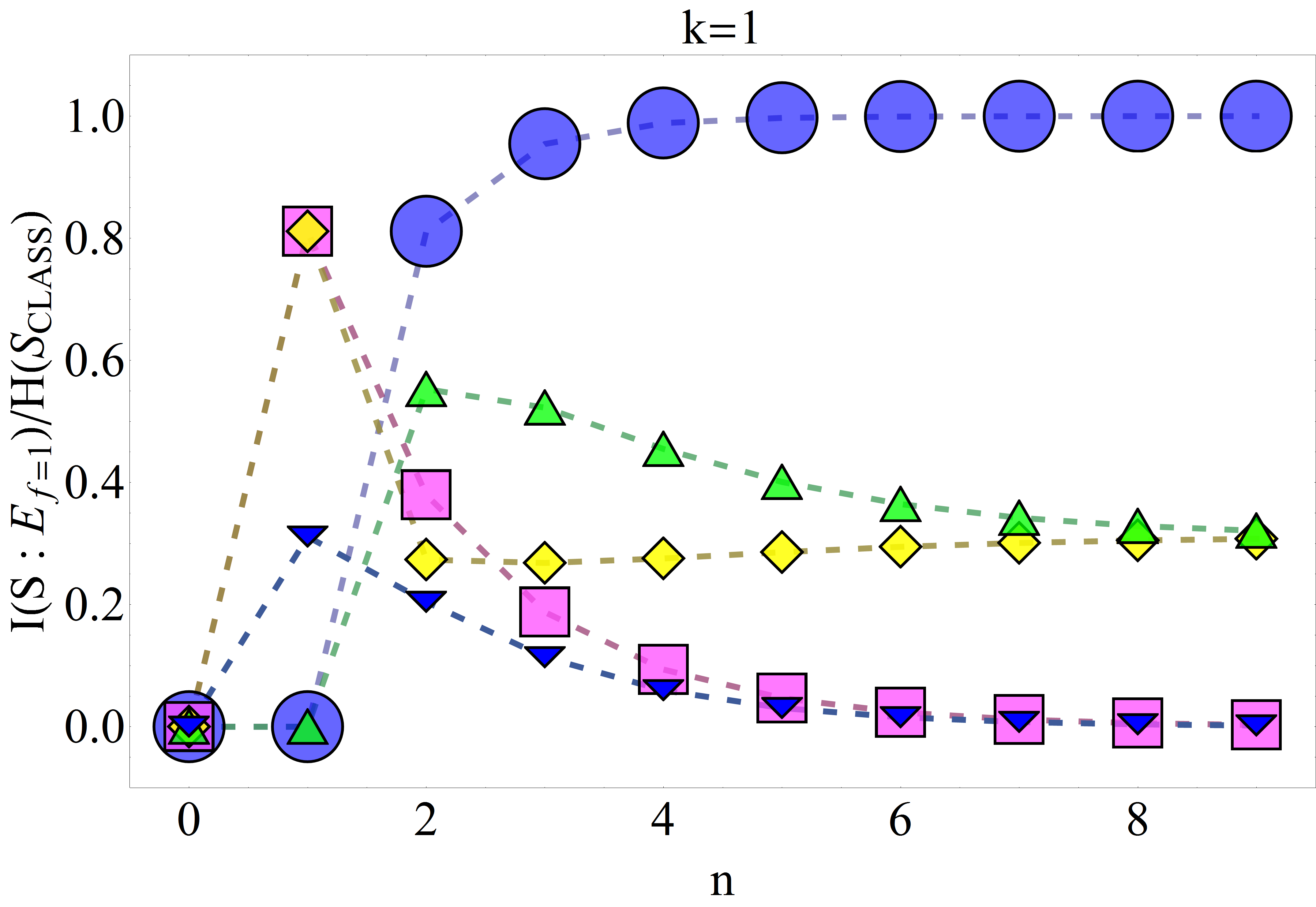}

\caption{$I\left(S:\, E_{L=n}\right)/H\left(S_{class}\right)$ vs $n$ after
random iterative $\widehat{u}_{j}^{\left(\phi=\pi/2\right)}$-evolution
of $\hat{\rho}_{SE}^{in}=\left|\Psi_{S}^{in}\right\rangle \left\langle \Psi_{S}^{in}\right|\otimes\hat{\rho}_{E}^{in}$ ($N\gg1$),
with a $k=1$ qubit $\left|\Psi_{S}^{in}\right\rangle =\protect\underset{m=0}{\protect\overset{1}{\sum}}a_{m}\left|m\right\rangle $
($\left|a_{m}\right|^{2}=2^{-1}\forall m$) and different
$\hat{\rho}_{E}^{in}$ (with $\geq2\left(n-1\right)$ $E$-bindings).
$I\left(S:\, E_{L=n}\right)/H\left(S_{class}\right)$ vs $n$ for
$\hat{\rho}_{E}^{in}=\left|0_{n}\right\rangle \left\langle 0_{n}\right|$
(blue, $\bullet$-dotted curve, $0$ $E$-bindings) is also displayed\label{Fig.4}.}
\end{figure}

As usual, $\hat{\rho}_{S}^{in}=|\Psi_{S,\,k}^{in}\rangle\langle
\Psi_{S,\,k}^{in}|$ is a pure $k$-qubit system. Numerous interesting conclusions can be obtained by looking at the
behavior of MI with respect to the number $n$ of $E$-qubits. For
instance, within the maximal attractor space we need at least $n\geq5$
$E$-qubits in order to see stable convergence of $I\left(S:\, E_{L=n}\right)/H\left(S_{class}\right)$,
as indicated by the blue, $\bullet$-dotted curve in Fig. \ref{Fig.4} associated with
\[\hat{\rho}_{SE}^{in}=\left|\Psi_{S,\,k=1}^{in}\right\rangle \left\langle \Psi_{S,\,k=1}^{in}\right|\otimes\hat{\rho}_{E}^{in},\]
where $\left|\Psi_{S,\,k=1}^{in}\right\rangle =\underset{m=0}{\overset{1}{\sum}}a_{m}\left|m\right\rangle $
($\left|a_{m}\right|^{2}=2^{-1}\forall m$) is a pure $k=1$-qubit system $S$, $n>k=1$ and $\hat{\rho}_{E}^{in}=\left|0_{n}\right\rangle \left\langle 0_{n}\right|$. 

However, for the minimal $\lambda=1$ attractor
subspace the same input state $\hat{\rho}_{SE}^{in}$ leads for $k=1$
to the output state in (\ref{Gl J1}) of Appendix \ref{AAD2-1} with non-zero
eigenvalues 
\begin{align}
\label{Gl 4.4}
\begin{split}
\lambda_{1}^{SE}&=\left|a_{1}\right|^{2}\cdot2^{-L}\,\,\left(2^{L}-1\,\rm{times}\right)\\
\lambda_{2/3}^{SE}&=\left(2^{-1}\left|a_{0}\right|^{2}+\left|a_{1}\right|^{2}\cdot2^{-L+1}\right)\pm \\
&\sqrt{2^{-2}\left|a_{0}\right|^{4}+\left|a_{1}\right|^{4}\cdot2^{-2\left(L+1\right)}-\epsilon_{L}}\\
\lambda_{1}^{E}&=\left|a_{0}\right|^{2}+\left|a_{1}\right|^{2}\cdot2^{-L}\\
\lambda_{2}^{E}&=\left|a_{1}\right|^{2}\cdot2^{-L}\,\,\left(2^{L}-1\,\rm{times}\right)\\
\lambda_{1/2}^{S}&=\frac{1}{2}\pm\sqrt{\frac{1}{4}-\left|a_{0}\right|^{2}\left|a_{1}\right|^{2}\left(1-2^{-2n}\right)}
\end{split}
\end{align}
(valid $\forall\,1\leq L\leq n$), where
\[\epsilon_{L}:=2^{-2n+L}\left(2^{2\left(n-L\right)-1}-1\right)\left|a_{0}\right|^{2}\left|a_{1}\right|^{2}.\]
The PIP of (\ref{Gl 4.4}) is given by the yellow, $\blacklozenge$-dotted curve in
Fig. \ref{Fig.6}. Apparently, the absence of the $\lambda=-1$ attractor
subspace is crucial for the appearence of Quantum Darwinism in case
of $\hat{\rho}_{E}^{in}=\left|0_{n}\right\rangle \left\langle 0_{n}\right|$.

On the other hand, the $\circ$-dotted and the $\square$-dotted curve
in Fig. \ref{Fig.6} also demonstrate what happens within the minimal
$\lambda=1$ attractor subspace for the output state in (\ref{Gl J1})
of Appendix \ref{AAD2-1} with $k\geq2$ $S$-qubits: since in the limit $n\gg k>1$
(\ref{Gl J1}) of Appendix \ref{AAD2-1} leads to the same form (\ref{Gl 4.1})
as (\ref{Gl I1}) of Appendix \ref{AAD1-1}, we see that with increasing number
$k$ of $S$-qubits (i.e. in the limit $n\sim k\gg1$) $I\left(S:\, E_{L=n}\right)/H\left(S_{class}\right)$
from (\ref{Gl J1}) of Appendix \ref{AAD2-1} will also behave (with $\left|a_{i}\right|^{2}=2^{-k}\,\forall i\in\left\{ 0,\,...,\,2^{k}-1\right\} $)
as 
\[
I\left(S:\, E_{L=n\sim k\gg1}\right)/H\left(S_{class}\right)\sim2^{-k}.
\]
Accordingly, one also has (again with equal $S$-probability distribution $\left|a_{i}\right|^{2}=2^{-k}\,\forall i\in\left\{ 0,\,...,\,2^{k}-1\right\} $)
\[
I\left(S:\, E_{L=n\gg k\gg1}\right)/H\left(S_{class}\right)=0.
\]

In Fig. \ref{Fig.4} we also see what happens with MI if $\hat{\rho}_{E}^{in}$,
such as $\hat{\rho}_{E}^{in}=\left|1_{n}\right\rangle \left\langle 1_{n}\right|$,
contains only $E$-registry states that do not participate within
a given, in this case minimal $\lambda=1$ attractor subspace: $I\left(S:\, E_{L=n}\right)/H\left(S_{class}\right)$
(red, $\blacksquare$-dotted curve) tends to zero in the limit $n\gg k$.
This can be easily explain by taking into account the fact that $\hat{\rho}_{SE_{L=n}}^{out}$
from (\ref{Gl J2}) in Appendix \ref{AAD2-1} acquires for a $k=1$ qubit $S$
the form
\begin{equation}
\underset{n\gg k}{\lim}\hat{\rho}_{SE_{L=n}}^{out}=2^{-n}\cdot\sum_{m=0}^{1}\left|a_{m}\right|^{2}\left|m\right\rangle \left\langle m\right|\otimes\widehat{I}_{n},\label{Gl 4.4.1}
\end{equation}
in the limit $n\gg k$, yielding \[H\left(S:\, E_{L=n\gg k}\right)=H\left(E_{L=n\gg k}\right)+H\left(S_{class}\right).\]
Thus, $\hat{\rho}_{E}^{in}$ that are not contained in (``recognized''
by) a minimal $\lambda=1$ attractor subspace do not contribute to
$I\left(S:\, E_{L}\right)/H\left(S_{class}\right)$ in the limit $n\gg k$.

\paragraph{II) $\hat{\rho}_{SE}^{in}=|\Psi_{S,\,k=1}^{in}\rangle\langle
\Psi_{S,\,k=1}^{in}|\otimes\ensuremath{\frac{1}{2}\left(\left|0_{n}\right\rangle \left\langle 0_{n}\right|+\left|1_{n}\right\rangle \left\langle 1_{n}\right|\right)}$
}

~

From Fig. \ref{Fig.4} (yellow, $\blacklozenge$-dotted curve) we
also conclude that this type of $\hat{\rho}_{SE}^{in}$ never leads
within the minimal $\lambda=1$ attractor subspace to \[I\left(S:\, E_{L=n}\right)/H\left(S_{class}\right)=1\]
in the limit $n\gg k$, since in this case the corresponding $\hat{\rho}_{SE}^{out}$
from (\ref{Gl J3}) in Appendix \ref{AAD2-1} acquires the form
\begin{equation}
\begin{array}{c}
\underset{n\gg k}{\lim}\hat{\rho}_{SE_{L=n}}^{out}=\left|a_{0}\right|^{2}\left|0\right\rangle \left\langle 0\right|\otimes\Biggl\{\widehat{I}_{n}2^{-n-1}\\
+\frac{1}{2}\left|0_{n}\right\rangle \left\langle 0_{n}\right|\Biggr\}+\left|a_{1}\right|^{2}\left|1\right\rangle \left\langle 1\right|\otimes2^{-n}\cdot\widehat{I}_{n}
\end{array},\label{Gl 4.5}
\end{equation}
which always yields \[\underset{n\gg k}{\lim}H\left(S,\, E_{L=n}\right)>\underset{n\gg k}{\lim}H\left(E_{L=n}\right)\]
(as can be easily confirmed from the corresponding eigenspectra of
$\underset{n\gg k}{\lim}\hat{\rho}_{SE_{L=n}}^{out}$ and $\underset{n\gg k}{\lim}\hat{\rho}_{SE}^{out}$,
s. also (\ref{Gl 4.2}) below). This means that correlation terms
in (\ref{Gl J3}) from Appendix \ref{AAD2-1}, that we deliberately ignored
in (\ref{Gl 4.5}), force the MI $I\left(S:\, E_{L=n}\right)/H\left(S_{class}\right)$
to converge to \[I\left(S:\, E_{L=n}\right)/H\left(S_{class}\right)=0.3\]
in the limit $n\gg k$. The same occurs if we choose \[\left|\Psi_{E}^{in}\right\rangle =2^{-1/2}\cdot\left(\left|0_{n}\right\rangle +\left|1_{n}\right\rangle \right)\]
as an environmental input state (green, $\blacktriangle$-dotted curve
in Fig. \ref{Fig.4}), since its $\hat{\rho}_{SE}^{out}$ would acquire in the limit $n\gg k$ the
same form (\ref{Gl 4.5}). Thus, no Quantum Darwinism appears for
these types of $\hat{\rho}_{E}^{in}$.

\paragraph{III) $\hat{\rho}_{SE}^{in}=|\Psi_{S,\,k=1}^{in}\rangle\langle
\Psi_{S,\,k=1}^{in}|\otimes2^{-n}\hat{I}_{n}$
}

~

As in (\ref{Gl 4.3}) below, $\hat{\rho}_{SE}^{out}$ from (\ref{Gl J4})
of Appendix \ref{AAD2-1} leads in the limit $n\gg k$ to \[H\left(S,\, E_{L=n\gg k}\right)=H\left(S_{class}\right)+H\left(E_{L=n\gg k}\right)\],
i.e. completely mixed $\hat{\rho}_{E}^{in}$ leads also within the
minimal $\lambda=1$ attractor subspace to the MI-value \[I\left(S:\, E_{L=n\gg k}\right)/H\left(S_{class}\right)=0\]
(s. Fig. \ref{Fig.4}, blue, $\blacktriangledown$-dotted curve).

\paragraph{IV) $\left|\Psi_{SE}^{in}\right\rangle =a\left|0_{k=1}\right\rangle \otimes\left|s_{1}^{L=n}\right\rangle +b\left|1_{k=1}\right\rangle \otimes\left|s_{2}^{L=n}\right\rangle $
}

~

The output state $\hat{\rho}_{SE}^{out}$ from (\ref{Gl J5}) of Appendix \ref{AAD2-1},
emerging from the random unitary evolution of this entangled, pure input state $\hat{\rho}_{SE}^{in}$,
and its eigenspectrum indicate that the relation \[
\ensuremath{H\left(S,\, E_{L}\right)-H\left(E_{L}\right)=\frac{1}{2}+}\ensuremath{\left(1-x_{L}\right)\log_{2}\left(1-x_{L}\right)>0,}
\]
where \[x_{L}:=2^{n-L-1}\cdot\left(2^{n}-1\right)^{-1},\] holds for all $0<L<n$. In other words, the corresponding PIP has the
same behavior as displayed by the blue, $\blacktriangledown$-dotted
curve in Fig. \ref{Fig.5}. Therefore, without the $\lambda=-1$ attractor
subspace the minimal $\lambda=1$ attractor subspace does not suffice
to ensure that Quantum Darwinism appears, as is the case with the
maximal attractor space discussed in subsection \ref{A3.3} below.

\subsection{Short time limit of the random unitary evolution\label{A3.2}}

~~~Before looking at the analytic structure of the corresponding
maximal attractor space we discuss whether one may interpret Zurek's
qubit model of Quantum Darwinism as the short time limit (corresponding
to the small number $N$ of iterations) of the random unitary evolution
involving pure decoherence. 

\begin{figure}[H]
\center\includegraphics[width=0.48\textwidth,natwidth=610,natheight=642]{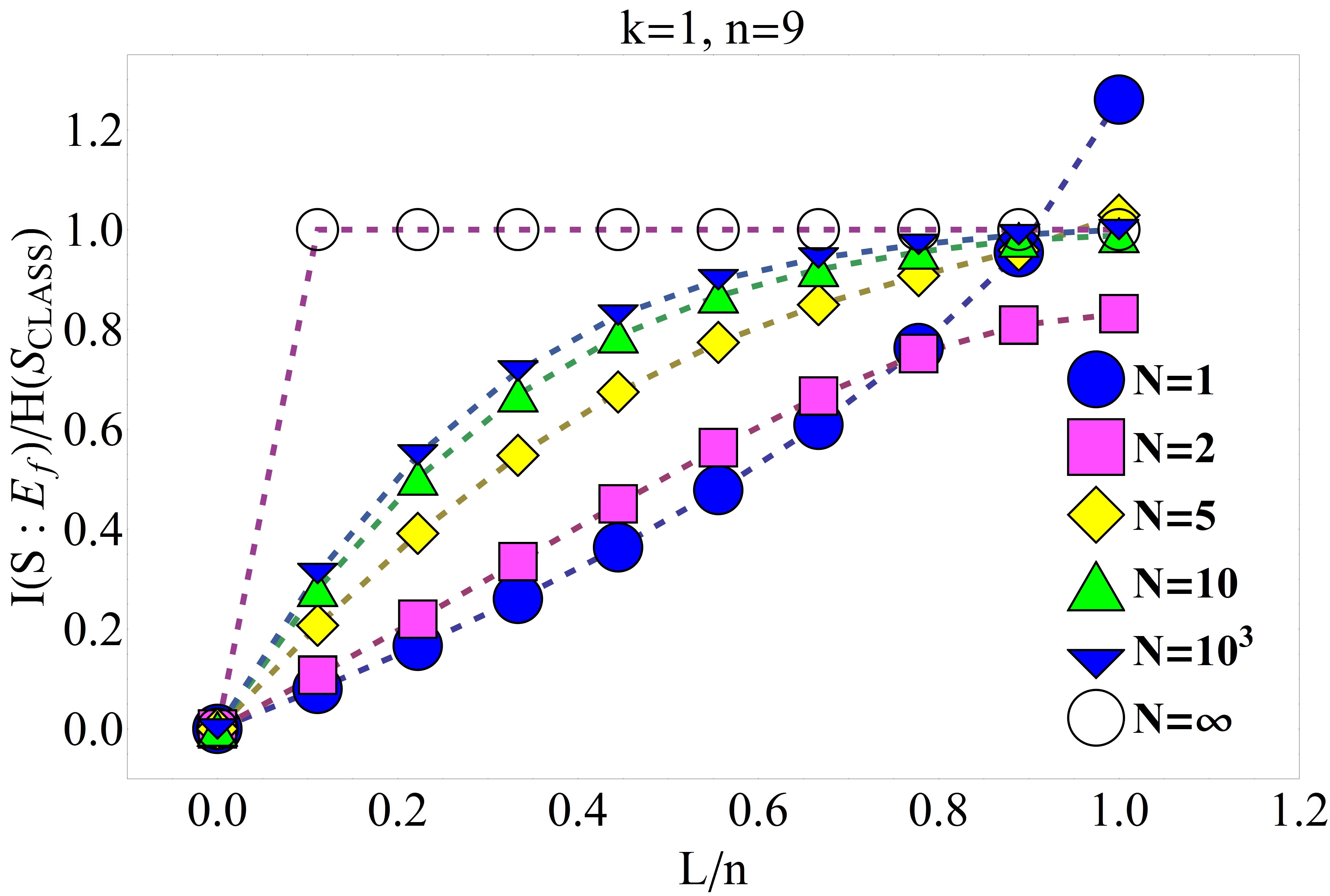}

\caption{PIP of simulated, random unitarily CNOT-evolved MI vs $0<f\leq1$ for $\hat{\rho}_{SE}^{in}=\hat{\rho}_{S}^{in}\otimes\hat{\rho}_{E}^{in}$
in (\ref{Gl 3.5}) and $\hat{\rho}_{SE}^{out}$ from (\ref{Gl 2.1})-(\ref{Gl 2.5}).
For $N=\infty$ s. subsection \ref{A3.3.2}. \label{Fig.5}}
\end{figure}

Within the random unitary operation-formalism we obtain another type
of PIP-behavior: inserting $\hat{\rho}_{SE}^{in}$ from Fig. \ref{Fig.1}
into (\ref{Gl 2.5}) we obtain for pure decoherence, with %\[
\begin{align*}
\begin{split}
\ensuremath{\left|a\right|^{2}&=\left|b\right|^{2}=1/2}\\
\ensuremath{p_{e}&=1/\left|M\right|}\,\forall e,
\end{split}
\end{align*}
%\] 
after $N\gg1$ iterations the PIP in
Fig. \ref{Fig.5}, which suggests that Zurek\textquoteright{}s Quantum
Darwinistic-\textquoteright{}plateau\textquoteright{} \cite{key-3}
appears only in the limit $N\rightarrow\infty$ (we will obtain this
asymptotic limit $N\rightarrow\infty$ of the random unitary evolution
analytically in subsection \ref{A3.3}). Thus, Zurek\textquoteright{}s
qubit model of Quantum Darwinism does not appear as the short-time
limit (small $N$-values, e.g. $N\leq10$) of our random unitary evolution
model with pure decoherence.

\subsection{Maximal attractor space\label{A3.3}}

~~~When dealing with Koenig-IDs \cite{key-6} we always obtain
attractor (sub-)spaces with maximal dimension $d^{\lambda}$ (determined
by (\ref{Gl G2})-(\ref{Gl G3}) in Appendix \ref{AAC}), since in such IDs
$E$-qubits are not allowed to interact with each other. Therefore,
we turn our attention in the following subsections to the description
of analytical attractor space structures associated with Koenig-IDs
and determined in Appendix \ref{AAC}.
\subsubsection{State structure of the attractor space\label{A3.3.1}}

~

From Appendix \ref{AAC} we know that for the random unitary evolution the
attractor space consists of two subspaces (\ref{Gl G4}) and (\ref{Gl G5})
(Appendix \ref{AAC1.2}) associated with eigenvalues $\lambda=1$ and $\lambda=-1$ of (\ref{Gl 2.6}),
respectively.

The main (largest) part of the attractor states $\widehat{X}_{\lambda=1,i}$
can be attributed to the $|0_{k}\rangle\langle0_{k}|$-subspace of
system $S$, since the $\lambda=1$-attractor subspace describes the
impact of pure decoherence on system $S$ during the iterative evolution
(\ref{Gl 2.5}) of $\hat{\rho}_{SE}^{in}$. However, in order to realize
the physical significance of the $\lambda=-1$-attractor subspace
we will discuss in the following subsection the random unitary evolution
of some of the $\hat{\rho}_{SE}^{in}$ from Tab. \ref{Tab. A1} that
have already been studied in the course of Zurek's evolution in section
\ref{A2}.

\subsubsection{Results of the CNOT-evolution\label{A3.3.2}}

~

Now we look at the random unitary CNOT-evolution from the analytical
point of view by utilizing the attractor space structure from subsection
\ref{A3.3.1} and concentrating on the following input states $\hat{\rho}_{SE}^{in}$
(with $n\gg k\geq1$) :

\paragraph{I) $\hat{\rho}_{SE}^{in}=|\Psi_{S,\,k}^{in}\rangle\langle
\Psi_{S,\,k}^{in}|\otimes\hat{\rho}_{E}^{in}$,
$\hat{\rho}_{E}^{in}=\left|y\right\rangle \left\langle y\right|$, $\ensuremath{y\in\left\{ 0_{n},\,1_{n}\right\} }$
}

~

Decomposing $\hat{\rho}_{SE}^{in}$ for $n\gg k\in\left\{ 1,\,2,\,3\right\} $
$S$-qubits by means of (\ref{Gl 2.7}) and $\widehat{X}_{\lambda,i}$
from (\ref{Gl G4})-(\ref{Gl G5}) of Appendix \ref{AAC1.2} that are already Gram-Schmidt orthonormalized,
we obtain the CNOT-asymptotically evolved $\hat{\rho}_{SE}^{out}$
displayed in (\ref{Gl I1}) of Appendix \ref{AAD1-1}. The corresponding PIP
obtained from $\hat{\rho}_{SE}^{out}$ in (\ref{Gl I1}) of Appendix \ref{AAD1-1}
for $k\in\left\{ 1,\,2,\,3\right\} $ $S$-qubits is displayed in
Fig. \ref{Fig.6} below.

Fig. \ref{Fig.6} demonstrates that within the random unitary operations
model Quantum Darwinism appears only for $k=1$ pure $\hat{\rho}_{S}^{in}$
even if we set as an environmental input state $\hat{\rho}_{E}^{in}=\left|y\right\rangle \left\langle y\right|$ for all $ y\in\left\{ 0,\,...,\,2^{n}-1\right\} $ and
with mutually non-interacting $E$-qubits, whereas for $n\sim k\gg1$
the maximal $I\left(S:\, E_{L=n}\right)$-value that can be achieved
after enclosing the entire environment $E$ behaves as 
\[
\underset{n\sim k\gg1}{\lim}I\left(S:\, E_{L=n}\right)/H\left(S_{class}\right)\sim2^{-k}.
\]
This follows from (\ref{Gl I1}) of Appendix \ref{AAD1-1} which, with (without
loss of generality) $\left|a_{i}\right|^{2}=2^{-k}\,\forall i\in\left\{ 0,\,...,\,2^{k}-1\right\} $
and for $k>1$, acquires in the limit $n\gg1$ the form

\begin{equation}
\begin{array}{c}
\underset{n\gg1}{\lim}\hat{\rho}_{SE_{L=n}}^{out}=\left|a_{0}\right|^{2}\left|0_{k}\right\rangle \left\langle 0_{k}\right|\otimes\left|0_{n}\right\rangle \left\langle 0_{n}\right|\\
+\underset{m=1}{\overset{2^{k}-1}{\sum}}\left|a_{m}\right|^{2}\left|m\right\rangle \left\langle m\right|\otimes2^{-n}\hat{I}_{n}.
\end{array}\label{Gl 4.1}
\end{equation}
(\ref{Gl 4.1}) leads to $H\left(S\right)\approx H\left(S_{class}\right)=k$
(with a decoherence factor $0\leq r=\left\langle s_{1}^{n}\right|\hat{\rho}_{E}^{in}\left|s_{1}^{n}\right\rangle =2^{-n}\leq1$),
and non-zero eigenvalues
%\[
\begin{align*}
\begin{split}
\lambda_{1}^{SE}&=\left|a_{0}\right|^{2}=2^{-k}\,\,\left(1\,\rm{times}\right)\\
\lambda_{2}^{SE}&=\left|a_{m}\right|^{2}2^{-n}=2^{-\left(k+n\right)}\,\,\left(2^{n}\left[2^{k}-1\right]\,\rm{times}\right),
\end{split}
\end{align*}
%\]
 yielding (for fixed $n$ and increasing $k$)
\[
\underset{n\sim k\gg1}{\lim}H\left(S,\, E_{L=n}\right)=2k+k\cdot2^{-k}.
\]
Accordingly, the eigenvalues of $\underset{n\gg1}{\lim}\hat{\rho}_{E_{L=n}}^{out}$
from (\ref{Gl 4.1}),
%\[
\begin{align*}
\begin{split}
\lambda_{1}^{E}&=\left|a_{0}\right|^{2}+2^{-n}\left(1-\left|a_{0}\right|^{2}\right)\,\,\left(1\rm{times}\right)\\
\lambda_{2}^{E}&=2^{-n}\left(1-\left|a_{0}\right|^{2}\right)\,\,\left(\left[2^{n}-1\right]\rm{times}\right),
\end{split}
\end{align*}
%\]
 lead (as in Fig. \ref{Fig.6}) to 
%\[
\begin{align*}
\begin{split}
&\underset{n\sim k\gg1}{\lim}H\left(E_{L=n}\right)=k+k\cdot2^{1-k}\\
&\underset{n\sim k\gg1}{\lim}I\left(S:\, E_{L=n}\right)/H\left(S_{class}\right)=2^{-k}.
\end{split}
\end{align*}
%\]

\begin{figure}[H]
\includegraphics[width=0.48\textwidth,natwidth=610,natheight=642]{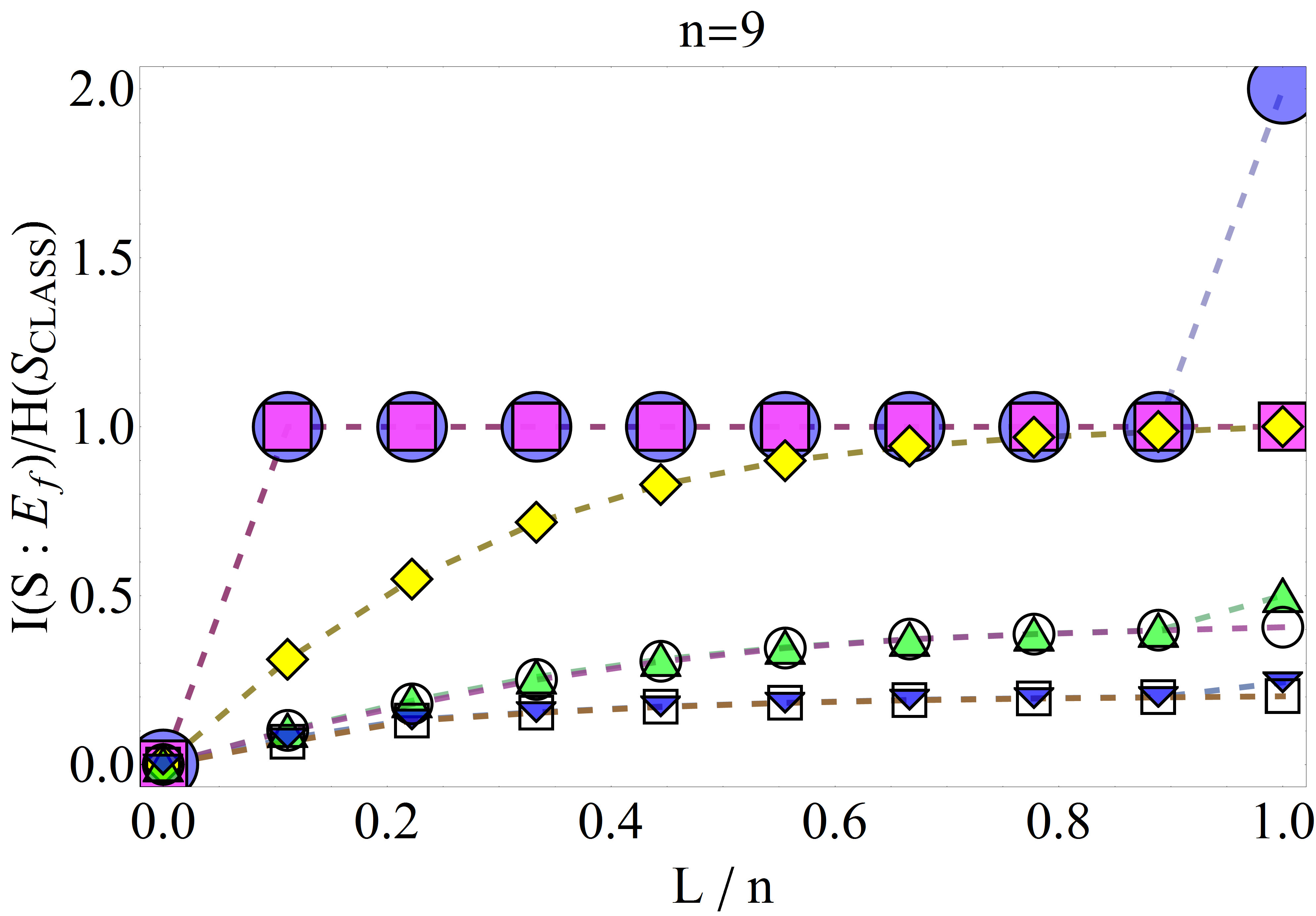}

\caption{PIP after random iterative $\widehat{u}_{j}^{\left(\phi=\pi/2\right)}$-evolution
of $\hat{\rho}_{SE}^{in}=\left|\Psi_{S}^{in}\right\rangle \left\langle \Psi_{S}^{in}\right|\otimes\left|0_{n=9}\right\rangle \left\langle 0_{n=9}\right|$ ($N\gg1$),
with a $k$ qubit $\left|\Psi_{S}^{in}\right\rangle =\protect\underset{m=0}{\protect\overset{2^{k}-1}{\sum}}a_{m}\left|m\right\rangle $
($\left|a_{m}\right|^{2}=2^{-k}\forall m$), without ($k=1$, $\blacksquare$-dotted
curve; $k=2$, $\blacktriangle$-dotted curve; $k=3$, $\blacktriangledown$-dotted
curve) and with $\geq1$ interaction bindings ($k=1$, $\blacklozenge$-dotted
curve; $k=2$, $\circ$-dotted curve; $k=3$, $\square$-dotted curve)
between $E$-qubits. The corresponding PIP of Zurek's model ($\bullet$-dotted
curve) is also displayed.\label{Fig.6}}
\end{figure}

Even worse: if we choose $k$ and $n$ sufficiently high, such as
$n\gg k\gg1$, (\ref{Gl 4.1}) yields (again with an $S$-probability distribution $\left|a_{i}\right|^{2}=2^{-k}\,\forall i\in\left\{ 0,\,...,\,2^{k}-1\right\} $)
%\[
\begin{align*}
\begin{split}
&\underset{n\gg k\gg1}{\lim}H\left(S,\, E_{L=n}\right)=k+n+k\cdot2^{-k}\\
&\underset{n\gg k\gg1}{\lim}H\left(E_{L=n}\right)=n+k\cdot2^{-k}\\
&\underset{n\gg k\gg1}{\lim}I\left(S:\, E_{L=n}\right)/H\left(S_{class}\right)=0.
\end{split}
\end{align*}
%\]
This is in conflict with the expectation of Zurek's CNOT-evolution
model, which predicts the appearence of the MI-'plateau' $\forall k\geq1$.
Apparently, the random unitary evolution model suggests that in order
to store $H\left(S_{class}\right)$ into environment $E$ efficiently
one needs environments consisting of qudit-cells ($2^{k}$-level systems).
This conjecture is also supported by (\ref{Gl E6}) in Appendix \ref{AAB}, which indicates
that for Quantum Darwinism to appear w.r.t. $k>1$ one needs $2^{k}$
symmetry states. Unfortunately, the qubit-qubit $\widehat{u}_{j}^{\left(\phi\right)}$-transformation
in (\ref{Gl 2.1})-(\ref{Gl 2.2}) (and thus also the CNOT) offers
only two symmetry states $\left\{ \left|s_{c_{1}}\right\rangle ,\,\left|s_{c_{2}}\right\rangle \right\} $.
Therefore, one would require a qubit-qudit version of (\ref{Gl 2.1})-(\ref{Gl 2.2})
in order to see Quantum Darwinism.

On the other hand, for $k=1$ $\hat{\rho}_{SE_{L}}^{out}$ and $\hat{\rho}_{E_{L}}^{out}$
from (\ref{Gl I1}) in Appendix \ref{AAD1-1} have $\forall\left(1\leq L\leq n\right)$ identical
non-zero eigenvalues
%\[
\begin{align*}
\begin{split}
\lambda_{1}^{\left(S\right)E}&=\left|a_{1}\right|^{2}2^{-L}\,\,(\left[2^{L}-1\right]\,\rm{times})\\
\lambda_{2}^{\left(S\right)E}&=2^{-1}\left|a_{0}\right|^{2}+2^{-\left(L+1\right)}\left|a_{1}\right|^{2}+\\
&\sqrt{\left(2^{-1}\left|a_{0}\right|^{2}+\left|a_{1}\right|^{2}2^{-\left(L+1\right)}\right)^{2}-g_{N}}\\
&=\left|a_{0}\right|^{2}+2^{-L}\left|a_{1}\right|^{2}\,\,(1\,\rm{times}),
\end{split}
\end{align*}
%\]
 with \[g_{N}:=\left|a_{0}\right|^{2}\left|a_{1}\right|^{2}2^{-2n+L}\left[\underset{:=0}{\underbrace{1-\left(-1\right)^{2N}}}\right],\]
due to the $\lambda=-1$ attractor subspace and its contributions
in $\lambda_{2}^{\left(S\right)E}$ proportional to $\left(-1\right)^{2N}$
and characterized by the iteration number $N$ from (\ref{Gl 2.7}).
Again, $H\left(S\right)\approx H\left(S_{class}\right)$ for $n\gg k$
due to eigenvalues 
\[
\lambda_{1/2}^{S}=1/2\pm\sqrt{1/4-\left(1-2^{-2n}\left[1+\left(-1\right)^{N}\right]^{2}\right)\left|a_{0}\right|^{2}\left|a_{1}\right|^{2}}
\]
 of $\hat{\rho}_{S}^{out}$ from (\ref{Gl I1}) in Appendix \ref{AAD1-1}.

\paragraph{II) $\hat{\rho}_{SE}^{in}=|\Psi_{S,\,k=1}^{in}\rangle\langle
\Psi_{S,\,k=1}^{in}|\otimes\ensuremath{\frac{1}{2}\left(\left|0_{n}\right\rangle \left\langle 0_{n}\right|+\left|1_{n}\right\rangle \left\langle 1_{n}\right|\right)}$}

~

\begin{figure}[H]
\includegraphics[width=0.48\textwidth,natwidth=610,natheight=642]{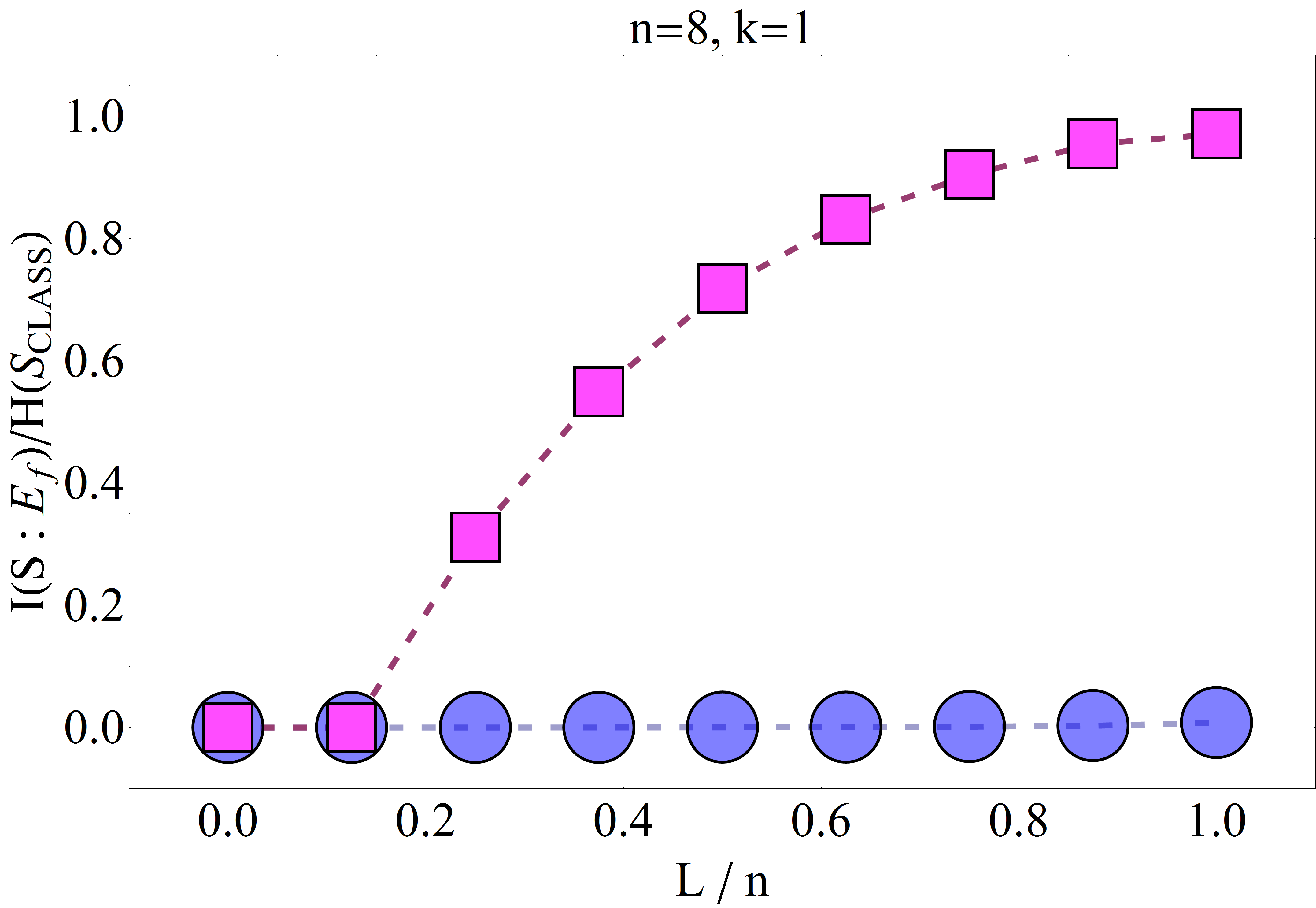}

\caption{PIP after random iterative $\widehat{u}_{j}^{\left(\phi=\pi/2\right)}$-evolution
of $\hat{\rho}_{SE}^{in}=\left|\Psi_{S}^{in}\right\rangle \left\langle \Psi_{S}^{in}\right|\otimes\hat{\rho}_{E}^{in}$ ($N\gg1$),
with a pure $k=1$ qubit $\left|\Psi_{S}^{in}\right\rangle =\protect\underset{m=0}{\protect\overset{1}{\sum}}a_{m}\left|m\right\rangle $
($\left|a_{m}\right|^{2}=2^{-1}\forall m$), $n=8$ and different
$\hat{\rho}_{E}^{in}$ ($0$ $E$-bindings), s. main text. \label{Fig.7}}
\end{figure}

The PIP obtained from a random unitarily $\widehat{u}_{j}^{\left(\phi=\pi/2\right)}$-evolved
$\hat{\rho}_{SE}^{out}$ in (\ref{Gl I3}) of Appendix \ref{AAD1-1} for $k=1$
$S$-qubit is displayed in Fig. \ref{Fig.7} below (red, $\blacksquare$-dotted
curve). We see that if $\hat{\rho}_{E}^{in}$ contains correlations
between $E$-registry states one is even not able to extract $H\left(S_{class}\right)$
after taking the entire $E$ into account when computing $I\left(S:\, E_{L}\right)$,
since according to Fig. \ref{Fig.7} \[I\left(S:\, E_{L=n}\right)/H\left(S_{class}\right)<1.\]

This can be easily explained by looking at the $n\gg k$ limit of
(\ref{Gl I3}) in Appendix \ref{AAD1-1} 
\begin{align}
\label{Gl 4.2}
\begin{split}
&\underset{n\gg k}{\lim}\hat{\rho}_{SE_{L=n}}^{out}=\left|a_{0}\right|^{2}\left|0_{k}\right\rangle \left\langle 0_{k}\right|\otimes\hat{\rho}_{E}^{in}\\
&+\left|a_{1}\right|^{2}\left|1\right\rangle \left\langle 1\right|\otimes2^{-n}\hat{I}_{n}\Rightarrow H\left(S\right)=H\left(S_{class}\right)\\
&\underset{n\gg k}{\lim}\hat{\rho}_{E_{L=n}}^{out}=\left|a_{0}\right|^{2}\hat{\rho}_{E}^{in}+\left|a_{1}\right|^{2}2^{-n}\hat{I}_{n}.
\end{split}
\end{align}
The (non-zero) eigenvalues
%\[
\begin{align*}
\begin{split}
\lambda_{1}^{SE}&=\left|a_{0}\right|^{2}2^{-1}\,\left(2\,\rm{times}\right)\\
\lambda_{2}^{SE}&=\left|a_{1}\right|^{2}2^{-n}\,\left(2^{n}\,\rm{times}\right)
\end{split}
\end{align*}
%\]
 (for $\underset{n\gg k}{\lim}\hat{\rho}_{SE_{L=n}}^{out}$), as well
as eigenvalues 
%\[
\begin{align*}
\begin{split}
\lambda_{1}^{E}&=\left|a_{0}\right|^{2}2^{-1}+\left|a_{1}\right|^{2}2^{-n}\,\left(2\,\rm{times}\right)\\
\lambda_{2}^{E}&=\left|a_{1}\right|^{2}2^{-n}\,\left(\left[2^{n}-2\right]\,\rm{times}\right)
\end{split}
\end{align*}
%\]
 (for $\underset{n\gg k}{\lim}\hat{\rho}_{E_{L=n}}^{out}$), yield
(for $0<\left|a_{0}\right|^{2}<1$)
%\[
\begin{align*}
\begin{split}
\underset{n\gg k}{\lim}H\left(S,\, E_{L=n}\right)&=H\left(S_{class}\right)+\left|a_{0}\right|^{2}\\
+\left|a_{1}\right|^{2}n&>\underset{n\gg k}{\lim}H\left(E_{L=n}\right),
\end{split}
\end{align*}
%\]
 since $\underset{n\gg k}{\lim}H\left(E_{L=n}\right)$ contains two
addends, 
%\[
\begin{align*}
\begin{split}
A_{1}:&=-\left|a_{1}\right|^{2}\log_{2}\frac{\left|a_{1}\right|^{2}}{2^{n}}+2^{1-n}\left|a_{1}\right|^{2}\log_{2}\frac{\left|a_{1}\right|^{2}}{2^{n}}\\
A_{2}:&=-\left(\left|a_{0}\right|^{2}+2^{1-n}\left|a_{1}\right|^{2}\right)\\
&\cdot\log_{2}\left(2^{-1}\left|a_{0}\right|^{2}+2^{-n}\left|a_{1}\right|^{2}\right),
\end{split}
\end{align*}
%\]
with
\begin{align*}
\begin{split}
A_{1}&<-\left|a_{1}\right|^{2}\log_{2}\frac{\left|a_{1}\right|^{2}}{2^{n}}\\ A_{2}&<-\left|a_{0}\right|^{2}\log_{2}\frac{\left|a_{0}\right|^{2}}{2}.
\end{split}
\end{align*}

In other words, if correlations between $E$-registry states persist
throughout the process of tracing out $E$-qubits from $\hat{\rho}_{SE_{L=n}}^{out}$,
(\ref{Gl 3.5.1}) will be violated and the MI-'plateau' disappears,
confirming the corresponding results obtained by means of Zurek's
model of Quantum Darwinism (s. also discussion from subsection \ref{A3.4}
below). Effectively the same PIP emerges if one starts the above random
unitary evolution with \[\left|\Psi_{E}^{in}\right\rangle =2^{-1/2}\left(\left|0_{n}\right\rangle +\left|1_{n}\right\rangle \right),\]
since contributions within the corresponding $\hat{\rho}_{SE}^{out}$
associated with non-classical correlation terms $\left|0_{n}\right\rangle \left\langle 1_{n}\right|$
and $\left|1_{n}\right\rangle \left\langle 0_{n}\right|$ also vanish
in the limit $n\gg k$ for all $\left(k\leq L<n\right)$.

\paragraph{III) $\hat{\rho}_{SE}^{in}=|\Psi_{S,\,k=1}^{in}\rangle\langle
\Psi_{S,\,k=1}^{in}|\otimes2^{-n}\hat{I}_{n}$
}

~

An extreme case for $\hat{\rho}_{E}^{in}$ containing classical correlations
between $E$-registry states is the totally mixed environmental $n$-qubit
input state which leads according to Fig. \ref{Fig.7} (blue, $\bullet$-dotted
curve) to \[\underset{n\gg k}{\lim}I\left(S:\, E_{L=n}\right)/H\left(S_{class}\right)\approx0.\]
This follows from the $n\gg k$ limit
\begin{align}
\label{Gl 4.3}
\begin{split}
&\underset{n\gg k}{\lim}\hat{\rho}_{SE_{L=n}}^{out}=\left(\left|a_{0}\right|^{2}\left|0_{k}\right\rangle \left\langle 0_{k}\right|+\left|a_{1}\right|^{2}\left|1\right\rangle \left\langle 1\right|\right)\otimes2^{-n}\hat{I}_{n}\\
&\Rightarrow H\left(S,\, E_{L=n\gg k}\right)=H\left(S_{class}\right)+H\left(E_{L=n\gg k}\right)
\end{split}
\end{align}
of (\ref{Gl I4}) in Appendix \ref{AAD1-1}. In other words, completely mixed
$\hat{\rho}_{E}^{in}$ are not suitable for efficiently storing $H\left(S_{class}\right)$
into $E$ with $f\leq k/n$.

\paragraph{IV) $\hat{\rho}_{SE}^{in}=|\Psi_{S,\,k=1}^{in}\rangle\langle
\Psi_{S,\,k=1}^{in}|\otimes\frac{\left(\left|0_{n}\right\rangle \left\langle 0_{n}\right|+\left|10_{n-1}\right\rangle \left\langle 10_{n-1}\right|\right)}{2}$
}

~

This type of $\hat{\rho}_{E}^{in}$ leads within the random unitary
model to $\hat{\rho}_{SE}^{out}$ in (\ref{Gl I5}) of Appendix \ref{AAD1-1},
demonstrating that within the random unitary model it is, as was the
case with Zurek's model, in principle important in which order one
traces out single $E$-qubits: if we trace out the first left $E$-qubit
in $\left|10_{n-1}\right\rangle \left\langle 10_{n-1}\right|$ for
a fixed $L$-value $L=L^{*}$ with $k\leq L^{*}<n$, $\hat{\rho}_{SE_{L*}}^{out}$
in (\ref{Gl I5}) of Appendix \ref{AAD1-1} would reduce to $\hat{\rho}_{SE_{L^{*}}}^{out}$
from (\ref{Gl I1}) of Appendix \ref{AAD1-1}, validating (\ref{Gl 3.5.1}) at least $\forall\left(k\leq L\leq L^{*}\right)$.

However, in general this is not what we demand from $\hat{\rho}_{SE}^{out}$
whose $E$ should allow complete reconstruction of the "classical" entropy $H\left(S_{class}\right)$
regardless of the order in which one decides to intercept environmental
fragments (qubits). This implies that among all possible combinations
(sums) of $E$-registry states only the pure (one) $E$-registry state
$\hat{\rho}_{E}^{in}=\left|y\right\rangle \left\langle y\right|$ in the standard computational basis (for all $y\in\left\{ 0,\,...\,2^{n}-1\right\} $) leads to Quantum Darwinism, both in Zurek's and the random unitary
model.

\paragraph{V) $\left|\Psi_{SE}^{in}\right\rangle =a\left|0_{k=1}\right\rangle \otimes\left|s_{1}^{L=n}\right\rangle +b\left|1_{k=1}\right\rangle \otimes\left|s_{2}^{L=n}\right\rangle $
}

~

This artificial $\hat{\rho}_{SE}^{in}$ entangles each $S$-pointer
state with one of the $\widehat{u}_{j}^{\left(\phi\right)}$-symmetry
states $\left\{ \left|s_{c_{1}}^{L}\right\rangle ,\,\left|s_{c_{2}}^{L}\right\rangle \right\} $,
validating (\ref{Gl 3.5.1}), according to Appendix \ref{AAB}, only for $c_{1}=c_{2}=1/2$,
which is why we obtain for the corresponding $\hat{\rho}_{SE}^{out}$
in (\ref{Gl I6}) of Appendix \ref{AAD1-1} exactly the same PIP as the one
displayed in Fig. \ref{Fig.1}: $\hat{\rho}_{SE}^{in}$ simply does
not change $\forall N\gg1$ due to invariance of $E$ towards $\widehat{u}_{j}^{\left(\phi=\pi/2\right)}$
and the fact that the $\lambda=-1$ attractor subspace in (\ref{Gl G5})
of Appendix \ref{AAC1.2} contributes to the random unitary evolution of $\hat{\rho}_{SE}^{in}$
for a $k=1$ qubit system $S$ only a phase $\left(-1\right)^{N}$-factor
within the $\left|0\right\rangle \left\langle 1\right|$- and $\left|1\right\rangle \left\langle 0\right|$-subspace
of system $S$. 

For $k>1$ one could obtain Quantum Darwinism according
to (\ref{Gl G4})-(\ref{Gl G5}) (Appendix \ref{AAC1.2}) only if one entangles two $S$-pointer
states $\left\{ \left|0_{k>1}\right\rangle ,\,\left|1_{k>1}\right\rangle \right\} $
with available CNOT-symmetry states $\left\{ \left|s_{1}^{L}\right\rangle ,\,\left|s_{2}^{L}\right\rangle \right\} $.
However, this would enable us to store only \[
\ensuremath{0<H\left(S\right)=H\left(S_{class}\right)\leq1},
\]
corresponding to a $k=1$ qubit system $S$. In order to store $H\left(S_{class}\right)$
of a $k>1$ qubit $S$ one needs $2^{k}$ symmetry states of $\widehat{U}_{ij}^{\left(\phi\right)}$
(s. (\ref{Gl 2.1}) above) with which one could entangle the $2^{k}$
$S$-pointer states $\left\{ \left|\pi_{i}\right\rangle \right\} \equiv\left\{ \left|i\right\rangle \right\} _{i=0}^{i=2^{k}-1}$,
otherwise if the number of $S$-pointer states exceeds the number
of available $\widehat{U}_{ij}^{\left(\phi\right)}$-symmetry states,
Quantum Darwinism disappears (s. (\ref{Gl E4})-(\ref{Gl E6}) in Appendix
\ref{AAB}).

\subsection{MI-Comparison: maximal vs minimal attractor space\label{A3.4}}

~~~Here we inquire which conclusions about the MI-behavior regarding
an increasing number of environmental qubit-qubit $\widehat{u}_{j}^{\left(\phi=\pi/2\right)}$-interactions
can be drawn simply by comparing the PIPs associated with both extrema
- the minimal and maximal attractor subspaces associated with an eigenvalue
$\lambda=1$ of (\ref{Gl 2.7}).

Indeed, many important conclusions about the behavior of the MI with
increasing number of $E$-qubit interactions in Fig. \ref{Fig.3}
can be drawn from a simple comparison between predictions obtained
by the random unitary evolution of $\hat{\rho}_{SE}^{in}$ in Fig.
\ref{Fig.1} from the point of view of the minimal and the maximal
$\lambda=1$ attractor subspaces (\ref{Gl G4}) and (\ref{Gl G8}) (Appendices  \ref{AAC1.2} and \ref{AAC2.2}),
respectively. For instance, looking at the PIP associated with the
maximal $\lambda=1$ attractor subspace (\ref{Gl G4}) alone (for
a $k\geq1$ qubit system $S$, s. Appendix \ref{AAC1.2}), which we obtain by ignoring all addends
in $\hat{\rho}_{SE}^{out}$ of (\ref{Gl I1}) from Appendix \ref{AAD1-1} proportional to $\left(-1\right)^{N}$,
we see that the MI behaves as in the PIP emerging from $\hat{\rho}_{SE}^{out}$
in (\ref{Gl J1}) of Appendix \ref{AAD2-1} evolved with respect to the minimal $\lambda=1$
attractor subspace. In other words, the PIP for $\hat{\rho}_{SE}^{out}$
(and a $k\geq1$ qubit system $S$) in (\ref{Gl I1}) of Appendix \ref{AAD1-1} without contributions
associated with the $\lambda=-1$ attractor subspace and the PIP obtained
from $\hat{\rho}_{SE}^{out}$ in (\ref{Gl J1}) (Appendix \ref{AAD2-1}) of the minimal $\lambda=1$
attractor subspace are exactly the same and are given by the $\blacklozenge$-,
$\circ$- and $\square$-dotted curves in Fig. \ref{Fig.6} (this
can also be confirmed numerically by iterating (\ref{Gl 2.5}) $N\gg1$
times). 

This means: $\hat{\rho}_{SE}^{out}$ from the $\lambda=1$-part
of (\ref{Gl I1}) in Appendix \ref{AAD1-1} and $\hat{\rho}_{SE}^{out}$ in (\ref{Gl J1}) of Appendix \ref{AAD2-1},
as can be readily confirmed, share the same non-zero eigenvalues (\ref{Gl 4.4}).
The presence of the $\lambda=-1$ attractor subspace (\ref{Gl G5}) from Appendix \ref{AAC1.2} in (\ref{Gl I1}) of Appendix \ref{AAD1-1} is essential for the appearence of Quantum Darwinism
(for a $k=1$ qubit system $S$) within the random unitary model.
Since the $\lambda=-1$ attractor subspace disappears from the attractor
space structure of a Koenig-like ID already after introducing a single
interaction arrow between two $E$-qubits (s. Appendix \ref{AAC}), the PIP
of a random unitarily evolved $\hat{\rho}_{SE}^{in}$ from Fig. \ref{Fig.1}
for environments $E$ containing one or more $\widehat{u}_{j}^{\left(\phi=\pi/2\right)}$-bindings
should be the same as the PIP of obtained from (\ref{Gl J1}) of Appendix \ref{AAD2-1} for
the minimal $\lambda=1$ attractor space, i.e. already a single interaction
between $E$-qubits in ID of Fig. \ref{Fig.3} destroys Quantum Darwinism
in the random unitary model.

For $\hat{\rho}_{E}^{in}\neq\left|y\right\rangle \left\langle y\right|\,\forall y\in\left\{ 0,\,...,\,2^{n}-1\right\} $
the contribution of the $\lambda=-1$ attractor subspace (\ref{Gl G5})
within the maximal attractor space (\ref{Gl G4})-(\ref{Gl G5}) from Appendix \ref{AAC1.2} to
the random unitary evolution of $\hat{\rho}_{SE}^{in}=\hat{\rho}_{S}^{in}\otimes\hat{\rho}_{E}^{in}$
and its MI-values is negligibly small in the limit $L=n\gg k$, whereas
the attractor subspace (\ref{Gl G4}) of Appendix \ref{AAC1.2} and its minimal version (\ref{Gl G8}) from Appendix \ref{AAC2.2}
dominate the asymptotic dynamics of $\hat{\rho}_{SE}^{in}$. 

Nevertheless, contributions from the $\lambda=-1$ attractor subspace do affect
outer-diagonal $S$-subspaces. For instance, (\ref{Gl I3}) in Appendix \ref{AAD1-1}
contains the most important part of the $\lambda=-1$ attractor subspace,
namely \[\left\{ \left|0_{L}\right\rangle \left\langle s_{2}^{L}\right|,\,\left|1_{L}\right\rangle \left\langle s_{2}^{L}\right|\right\} \]
(and their hermitean counterparts), within $S$-subspaces $\left|0\right\rangle \left\langle 1\right|$
and $\left|1\right\rangle \left\langle 0\right|$. When looking at
(\ref{Gl I3}) in Appendix \ref{AAD1-1} we see that these outer-diagonal $S$-subspaces
are associated with matrix entries \[\left[\left|0_{L}\right\rangle +\left|1_{L}\right\rangle \right]\left\langle \Psi_{E}^{combi}\right|\]
(and their hermitean counterpart, respectively), where 
\[
\left|\Psi_{E}^{combi}\right\rangle =\left|s_{1}^{L}\right\rangle +\left(-1\right)^{2n-L+N}\left|s_{2}^{L}\right\rangle 
\]
 distributes within the $\left|0_{L}\right\rangle $-th and $\left|1_{L}\right\rangle $-th
row (column) of (\ref{Gl I3}) in Appendix \ref{AAD1-1} $2^{L-1}$ complex-valued,
identical entries $c=2^{1-n}a_{0} a_{1}^{*}$ (alias its conjugate
counterparts). If we ask ourselves what is the ideal value $c_{ideal}$
of these $2^{L-1}$ identical entries in (\ref{Gl I3}) of Appendix \ref{AAD1-1}, distributed
within the $\left|0_{L}\right\rangle $-th and $\left|1_{L}\right\rangle $-th
row (column) in accord with $\left|\Psi_{E}^{combi}\right\rangle $,
for which the entropy-difference with respect to $\hat{\rho}_{SE}^{out}$ and $\hat{\rho}_{E}^{out}$, $H\left(S,\, E_{L=n}\right)-H\left(E_{L=n}\right)$
(with $n\gg k$), is minimal, we easily obtain \[c_{ideal}=2^{-n}a_{0} a_{1}^{*}\neq2^{1-n}a_{0} a_{1}^{*},\]
leading us for $\hat{\rho}_{SE}^{out}$ in (\ref{Gl I3}) of Appendix \ref{AAD1-1} to eigenvalues
\begin{equation}
\begin{array}{c}
\left.\begin{array}{c}
\lambda_{1}^{SE}=2^{-1}\left|a_{0}\right|^{2}\\
\lambda_{2}^{SE}=2^{1-L}\left|a_{1}\right|^{2}\left(2^{L-1}-1\,\textrm{times}\right)
\end{array}\right\} (\textrm{for}\,\hat{\rho}_{SE}^{out})\\
\left.\begin{array}{c}
\lambda_{1}^{E}=2^{-1}\left|a_{0}\right|^{2}+2^{1-L}\left|a_{1}\right|^{2}\,\left(2\,\textrm{times}\right)\\
\lambda_{2}^{E}=2^{1-L}\left|a_{1}\right|^{2}\,\left(2^{L-1}-2\,\textrm{times}\right)
\end{array}\right\} (\textrm{for}\,\hat{\rho}_{E}^{out}),
\end{array}\label{Gl 4.7}
\end{equation}
 that, in turn, yield $H\left(S,\, E_{L=n}\right)>H\left(E_{L=n}\right)$.
In other words, even in case of outer-diagonal $c$-entries in (\ref{Gl I3})
from Appendix \ref{AAD1-1} fixed as $c=c_{ideal}$ $H\left(S,\, E_{L=n}\right)$ would still
always exceed $H\left(E_{L=n}\right)$. 

The reason for this is connected with the following fact: for (\ref{Gl I1}) of Appendix \ref{AAD1-1},
emerging from the random unitary CNOT-evolution of $\hat{\rho}_{E}^{in}=\left|0_{n}\right\rangle \left\langle 0_{n}\right|$,
the diagonal value $\left|a_{0}\right|^{2}$ from the diagonal $S$-subspace
$\left|0\right\rangle \left\langle 0\right|$ in (\ref{Gl I1}) of Appendix \ref{AAD1-1} merges
with one of the diagonal values $2^{-L}\left|a_{1}\right|^{2}$ from
the diagonal $S$-subspace $\left|1\right\rangle \left\langle 1\right|$
after extracting $\hat{\rho}_{E}^{out}$ from $\hat{\rho}_{SE}^{out}$
and thus decreases $H\left(E_{L}\right)$ with respect to $H\left(S,\, E_{L}\right)$.
Fortunately, for this case $\left|\Psi_{E}^{combi}\right\rangle $
is the only combination that ca be made from two available CNOT-symmetry
states $\left\{ \left|s_{1}^{L}\right\rangle ,\,\left|s_{2}^{L}\right\rangle \right\} $
capable of reducing $H\left(S,\, E_{L}\right)$ such that $H\left(S,\, E_{L}\right)=H\left(E_{L}\right)\,\forall\left(1\leq L\leq n\right)$.
Unfortunately, in order to correct a higher number of overlapping
diagonal values between $S$-subspaces $\left|0\right\rangle \left\langle 0\right|$
and $\left|1\right\rangle \left\langle 1\right|$ within $\hat{\rho}_{E}^{out}$
(in (\ref{Gl I3}) of Appendix \ref{AAD1-1} there are two merging diagonal
values between $S$-subspaces $\left|0\right\rangle \left\langle 0\right|$ and
$\left|1\right\rangle \left\langle 1\right|$) one would also need
more than two symmetry states which is impossible for the CNOT transformation
and, in general, for the $\phi$-parameter family $\widehat{u}_{j}^{\left(\phi\right)}$
of transformations in (\ref{Gl 2.1})-(\ref{Gl 2.2}) (however, a
higher number of symmetry states is possible for a generalized, qudit-qudit
version of the CNOT-transformation). 

Therefore, $\hat{\rho}_{SE}^{in}=\hat{\rho}_{S}^{in}\otimes\hat{\rho}_{E}^{in}$
(with $\hat{\rho}_{S}^{in}$ being a pure $k=1$ qubit system $S$),
when being subject to CNOT-random unitary evolution leads in the asymptotic
limit $N\gg1$ of many iterations to Quantum Darwinism only if $\hat{\rho}_{E}^{in}=\left|y\right\rangle \left\langle y\right|\,\forall y\in\left\{ 0,\,...,\,2^{n}-1\right\} $,
otherwise, for $E$-input states $\hat{\rho}_{E}^{in}\neq\left|y\right\rangle \left\langle y\right|\,\forall y\in\left\{ 0,...,\,2^{n}-1\right\} $
the $\lambda=-1$ attractor subspace (\ref{Gl G5}) of Appendix \ref{AAC1.2} does not suffice to compensate all losses of $H\left(E_{L}\right)$ induced in $\hat{\rho}_{E}^{out}$
by overlapping diagonal entries within different diagonal $S$-subspaces.

Furthermore, by comparing the $\blacksquare$-dotted curve in Fig.
\ref{Fig.7} with the $\blacklozenge$-dotted curve in Fig. \ref{Fig.4}
we may conclude that the highest amount of asymptotic MI-values one
could achieve $\forall\left(k\leq L\leq n\right)$ is bounded from
above by $I\left(S:\, E_{L}\right)$ obtained from the maximal attractor
space (\ref{Gl G4})-(\ref{Gl G5}) of Appendix \ref{AAC1.2}.

\section{Summary and outlook\label{A4}}

~~~In this paper we studied the appearence of Quantum Darwinism
in the framework of the random unitary qubit model and compared the
corresponding results (Partial Information Plots of mutual information
between an open $k\geq1$ qubit system $S$ and its $n\gg k$ qubit
environment $E$) with respective predictions obtainable from Zurek's
qubit toy model. 

We found that the only $S$-$E$-input states $\hat{\rho}_{SE}^{in}$
which lead to Quantum Darwinism within the random unitary operations
model with maximal efficiency $f=f^{*}=k/n$, regardless of the order
in which one traces out single $E$-qubits, are the entangled input
state from equation (\ref{Gl 3.6}) and the product state $\hat{\rho}_{SE}^{in}=\hat{\rho}_{S}^{in}\otimes\hat{\rho}_{E}^{in}$,
with a pure $k=1$ qubit $\hat{\rho}_{S}^{in}$ and a pure one-registry
state $\hat{\rho}_{E}^{in}=\left|y\right\rangle \left\langle y\right|\,\forall y\in\left\{ 0,\,...,\,2^{n}-1\right\} $
of $n$ mutually non-interacting qubits in the standard computational
basis (Koenig-IDs). 

According to the random unitary operations model one is motivated
to conjecture that $\hat{\rho}_{SE}^{in}=\hat{\rho}_{S}^{in}\otimes\hat{\rho}_{E}^{in}$
with a pure $k>1$ qubit $\hat{\rho}_{S}^{in}$ allows efficient storage
of system's Shannon-entropy $H\left(S_{class}\right)$ into environment
$E$ only if $\hat{\rho}_{E}^{in}$ is given by a pure one registry
state $\hat{\rho}_{E}^{in}=\left|y'\right\rangle \left\langle y'\right|$
of mutually non-interacting $n$ qudits ($n$ $2^{k}$-level systems)
in the standard computational basis, with $y'\in\left\{ 0,\,...,\,2^{k\cdot n}-1\right\} $. 

This does not correspond to expectations arising from Zurek's qubit
model of Quantum Darwinism, which predicts the appearence of the mutual
information-'plateau' even for a $k>1$ qubit pure $\hat{\rho}_{S}^{in}$
and an $n$ qubit $\hat{\rho}_{E}^{in}$ within the aforementioned
$\hat{\rho}_{SE}^{in}=\hat{\rho}_{S}^{in}\otimes\hat{\rho}_{E}^{in}$,
indicating that Quantum Darwinism depends on the specific model on
which one bases his interpretations. Furthermore, the random unitary
model and Zurek's model of Quantum Darwinism must not be confused
with each other, since the latter does not correspond to the short
time limit (small iteration values $N$) of the former.

On the other hand, both in Zurek's and the random unitary model we
are able to confirm that correlations between qubit $E$-registry
states in $\hat{\rho}_{E}^{in}$, even if interactions between $E$-qubits
are absent, tend to suppress the appearence of the mutual information-'plateau'.
Also, the random unitary model indicates that already a single interaction
between $E$-qubits suppresses Quantum Darwinism.

If the Quantum Darwinistic description of the emergence of classical
$S$-states were correct, then Zurek's and the random unitary model
suggest that an open (observed) system $S$ of interest and its environment
$E$ must have started their evolution as a product state $\hat{\rho}_{SE}^{in}=\hat{\rho}_{S}^{in}\otimes\hat{\rho}_{E}^{in}$
with $\hat{\rho}_{S}^{in}$ denoting a pure $k\geq1$ qubit state
and $\hat{\rho}_{E}^{in}$ (denoting for instance the state of the
rest of the universe) given by a pure one-registry state of mutually
non-interacting $n$ qudits.

The above ``qudit-cell'' conjecture regarding environment $E$ of
the random unitary model could be tested by explicitly determining
the maximal attractor space between a $k>1$ qubit system $S$ and
its environment $E$ of mutually non-interacting $n$ qudits under
the impact of the generalized qubit-qudit version of the CNOT transformation
and focussing on the behavior of the mutual information within the
corresponding Partial Information Plot for such maximal attractor
space (Koenig-IDs). Furthermore, one could also ask what happens with
the efficiency of storing $H\left(S_{class}\right)$ into environment
$E$ if one introduces into the above random unitary evolution with
pure decoherence dissipative effects that would in general treat the
system $S$ in the interaction digraph of Fig. \ref{Fig.3} not only
as a control but also as a target, allowing $E$-qubits to react on
``impulses'' sent by $S$-qubits (paper in preparation).

\paragraph{Acknowledgements}

~

The author thanks G. Alber, J. Novotn$\textrm{\ensuremath{\acute{y}}}$
and J. Rennes for stimulating discussions. 

\paragraph{Author contribution statement}

~

The results of this paper were obtained by the author (N. Balaneskovi$\textrm{\textrm{\ensuremath{\acute{\textrm{c}}}}}$) in the framework of his PhD-research.

\appendix
%APPENDIX SECTION A

\twocolumn[{%
\section{List of exemplary input and output states in Zurek's model of Quantum Darwinism \label{AAA}}
In the present appendix we list all exemplary environmental input states in $\hat{\rho}_{SE}^{in}=\hat{\rho}_{S}^{in}\otimes\hat{\rho}_{E}^{in}$ and their output states $\hat{\rho}_{SE_{L}}^{out}$ discussed in the course of Zurek's qubit model of Quantum Darwinism in section \ref{A2}, Fig. \ref{Fig.2}.
\begin{table}[H] %\setlength\mathindent{0pt}
\begin{threeparttable}% Package ermöglicht das Anpassen der caption-Länge an die Tabellenlänge
\begin{tabular}{|l|l|} \hline  $\hat{\rho}_{E}^{in}$ & $\hat{\rho}_{SE_{L}}^{out}$ / entropies \tabularnewline \hline  \hline  1)$\begin{array}{c} \frac{1}{2}\left(\left|0_{n}\right\rangle \left\langle 0_{n}\right|+\left|0_{n-1}1\right\rangle \left\langle 0_{n-1}1\right|\right)\\ H\left(S\right)=H\left(S_{class}\right)\,\forall\, L\\ \bullet-\textrm{dotted\,\ curve} \end{array}$ & $\begin{array}{cc} \begin{array}{c} \left|a\right|^{2}\left|0\right\rangle \left\langle 0\right|\otimes\hat{\rho}_{E}^{in}\\ +\frac{1}{2}\left|b\right|^{2}\left|1\right\rangle \left\langle 1\right|\otimes\left(\left|1_{n}\right\rangle \left\langle 1_{n}\right|+\left|1_{n-1}0\right\rangle \left\langle 1_{n-1}0\right|\right)\\ +\frac{1}{2}ab^{*}\left|0\right\rangle \left\langle 1\right|\otimes\left(\left|0_{n}\right\rangle \left\langle 1_{n}\right|+\left|0_{n-1}1\right\rangle \left\langle 1_{n-1}0\right|\right)\\ +\frac{1}{2}a^{*}b\left|1\right\rangle \left\langle 0\right|\otimes\left(\left|1_{n}\right\rangle \left\langle 0_{n}\right|+\left|1_{n-1}0\right\rangle \left\langle 0_{n-1}1\right|\right) \end{array} & \left\{ \begin{array}{c} H_{SE}=H\left(S_{class}\right)+\delta_{L,n}\\ H_{E}=\left(1-\delta_{L,n}\right)H\left(S_{class}\right)\\ +\delta_{L,n}\left(1\leq L\leq n\right) \end{array}\right.\end{array}$\tabularnewline \hline  2)$\begin{array}{c} \frac{1}{2}\left(\left|0_{n}\right\rangle \left\langle 0_{n}\right|+\left|10_{n-1}\right\rangle \left\langle 10_{n-1}\right|\right)\\ H\left(S\right)=H\left(S_{class}\right)\,\forall\, L\\ \blacksquare-\textrm{dotted\,\ curve} \end{array}$ & $\begin{array}{cc} \begin{array}{c} \textrm{as\,\ in}\,\textrm{1),\,\ with}\\ \begin{array}{c} \left|0_{n-1}1\right\rangle \leftrightarrow\left|10_{n-1}\right\rangle \\ \left|1_{n-1}0\right\rangle \leftrightarrow\left|01_{n-1}\right\rangle  \end{array} \end{array} & \left\{ \begin{array}{c} H_{SE}=\left(1-\delta_{L,n}\right)H\left(S_{class}\right)+1\,\left(1\leq L\leq n\right)\\ H_{E}=1\,\textrm{for}\, L=1\\ H_{E}=\left(1-\delta_{L,n}\right)H\left(S_{class}\right)+H_{SE}\,\left(2\leq L\leq n\right) \end{array}\right.\end{array}$\tabularnewline \hline  3)$\begin{array}{c} \frac{1}{2}\left(\left|0_{n}\right\rangle \left\langle 0_{n}\right|+\left|1_{n}\right\rangle \left\langle 1_{n}\right|\right)\\ H\left(S\right)=H\left(S_{class}\right)\,\forall\, L\\ \blacklozenge-\textrm{dotted\,\ curve} \end{array}$ & $\begin{array}{cc} \begin{array}{c} \left(\left|a\right|^{2}\left|0\right\rangle \left\langle 0\right|+\left|b\right|^{2}\left|1\right\rangle \left\langle 1\right|\right)\otimes\hat{\rho}_{E}^{in}\\ +\frac{1}{2}\left(ab^{*}\left|0\right\rangle \left\langle 1\right|+a^{*}b\left|1\right\rangle \left\langle 0\right|\right)\\ \otimes\left(\left|0_{n}\right\rangle \left\langle 1_{n}\right|+\left|1_{n}\right\rangle \left\langle 0_{n}\right|\right) \end{array} & \left\{ \begin{array}{c} \left(1\leq L\leq n\right):\, H_{E}=1\\ H_{SE}=\left(1-\delta_{L,n}\right)H\left(S_{class}\right)+1 \end{array}\right.\end{array}$\tabularnewline \hline  4)$\begin{array}{c} 2^{-n}\hat{I}_{n}=\left(2^{-1}\hat{I}_{1}\right)^{\otimes n}\\ \left\{ \left|x\right\rangle ,\,\left|y\right\rangle \right\} \in\left\{ 0,\,...,\,2^{n}-1\right\} \\ H\left(S\right)=H\left(S_{class}\right)\,\forall\, L\\ \blacklozenge-\textrm{dotted\,\ curve} \end{array}$ & $\begin{array}{cc} \begin{array}{c} \left(\left|a\right|^{2}\left|0\right\rangle \left\langle 0\right|+\left|b\right|^{2}\left|1\right\rangle \left\langle 1\right|\right)\otimes\hat{\rho}_{E}^{in}\\ +2^{-n}ab^{*}\left|0\right\rangle \left\langle 1\right|\otimes\underset{x\neq y=0}{\overset{2^{n}-1}{\sum}}\left|x\right\rangle \left\langle y\right|\\ +2^{-n}a^{*}b\left|1\right\rangle \left\langle 0\right|\otimes\underset{x\neq y=0}{\overset{2^{n}-1}{\sum}}\left|y\right\rangle \left\langle x\right| \end{array} & \left\{ \begin{array}{c} \left(1\leq L\leq n\right):\, H_{E}=L\\ H_{SE}=\left(1-\delta_{L,n}\right)H\left(S_{class}\right)+L \end{array}\right.\end{array}$\tabularnewline \hline  5)$\begin{array}{c} \frac{1}{2}\left(\left|0_{n}\right\rangle \left\langle 0_{n}\right|+\left|1_{n-1}0\right\rangle \left\langle 1_{n-1}0\right|\right)\\ H\left(S\right)=H\left(S_{class}\right)\,\forall\, L\\ \blacktriangle-\textrm{dotted\,\ curve} \end{array}$ & $\begin{array}{cc} \begin{array}{c} \textrm{as\,\ in}\,\textrm{1),\,\ with}\\ \left|0_{n-1}1\right\rangle \leftrightarrow\left|1_{n-1}0\right\rangle  \end{array} & \left\{ \begin{array}{c} \left(1\leq L\leq n\right):\, H_{E}=1+\delta_{L,n}H\left(S_{class}\right)\\ H_{SE}=1+\left(1-\delta_{L,n}\right)H\left(S_{class}\right) \end{array}\right.\end{array}$\tabularnewline \hline  6)$\begin{array}{c} \left|\Psi_{E}^{in}\right\rangle \left\langle \Psi_{E}^{in}\right|\\ \left|\Psi_{E}^{in}\right\rangle =\frac{1}{\sqrt{2}}\left(\left|0_{n}\right\rangle +\left|1_{n}\right\rangle \right)\\ \blacktriangledown-\textrm{dotted\,\ curve} \end{array}$ & $\begin{array}{cc} \hat{\rho}_{S}^{in}\otimes\hat{\rho}_{E}^{in} & \left\{ \begin{array}{c} H\left(S\right)=0<\,\forall L>0\\ H_{E}=H_{SE}=1-\delta_{L,n}\,\left(1\leq L\leq n\right) \end{array}\right.\end{array}$\tabularnewline \hline  \end{tabular}
\caption{$\hat{\rho}_{SE_{L}}^{out}$ from Zurek's CNOT-evolution of $\hat{\rho}_{SE}^{in}=\hat{\rho}_{S}^{in}\otimes\hat{\rho}_{E}^{in}$ for different $\hat{\rho}_{E}^{in}$, s. Fig. \ref{Fig.2}. \label{Tab. A1}} 
\end{threeparttable}
\end{table}
}]

\section{Quantum Darwinism and eigenstates of (\ref{Gl 2.1})-(\ref{Gl 2.2})\label{AAB}}

~~~In this appendix we explain why the generalized $k>1$ qubit
version of (\ref{Gl 3.6}) does not lead to Quantum Darwinism. 

The $\phi$-parameter family $\widehat{u}_{j}^{\left(\phi\right)}$
of transformations in (\ref{Gl 2.1})-(\ref{Gl 2.2}) has eigenstates
$\left|s_{c_{1}}\right\rangle =\left(c_{1}\left|0\right\rangle +c_{2}\left|1\right\rangle \right)$
(eigenvalue $\lambda=1$) and $\left|s_{c_{2}}\right\rangle =\left(c_{2}\left|0\right\rangle -c_{1}\left|1\right\rangle \right)$
(eigenvalue $\lambda=-1$), with $\left\langle s_{c_{1}}|s_{c_{2}}\right\rangle =0$
and $c_{1}^{2}+c_{2}^{2}\overset{!}{=}1$ ($c_{1},\, c_{2}>0$) \cite{key-1,key-2,key-4}.
This allows us to parametrize \[
\ensuremath{c_{1}=\cos\left(\frac{\phi}{2}\right),\, c_{2}=\sin\left(\frac{\phi}{2}\right)}
\]
and thus fix $\phi$ within the range $\left(0\leq\phi\leq\pi\right)$.
By means of this $\phi$-parametrization we may generalize (\ref{Gl 3.6})
according to
\begin{equation}
\begin{array}{c}
\left|\Psi_{SE}^{out}\left(L=n\right)\right\rangle =a\left|0\right\rangle \otimes\left|s_{c_{1}}^{L}\right\rangle +b\left|1\right\rangle \otimes\left|s_{c_{2}}^{L}\right\rangle \\
\hat{\rho}_{E}^{out}\left(L=n\right)=\left|a\right|^{2}\left|s_{c_{1}}^{L}\right\rangle \left\langle s_{c_{1}}^{L}\right|+\left|b\right|^{2}\left|s_{c_{2}}^{L}\right\rangle \left\langle s_{c_{2}}^{L}\right|,
\end{array}\label{Gl E1}
\end{equation}
with $\left|s_{c_{1}}^{L}\right\rangle =\left|s_{c_{1}}\right\rangle ^{\otimes L},\,\left|s_{c_{2}}^{L}\right\rangle =\left|s_{c_{2}}\right\rangle ^{\otimes L},\,\left\langle s_{c_{1}}^{L}|s_{c_{2}}^{L}\right\rangle =0$.
For $L=n$ one would always obtain $H\left(S\right)=H\left(S_{class}\right)$
and $H\left(S,\, E_{L=n}\right)=0<H\left(E_{L=n}\right)$, since (\ref{Gl E1})
is a pure state, whereas the spectrum of $\hat{\rho}_{E}^{out}\left(L=n\right)$
would, for simplicity for $L=n=1$, contain the non-vanishing eigenvalues
\[\lambda_{1/2}^{E\left(L=n=1\right)}=\frac{1}{2}\pm\sqrt{\frac{1}{4}-4c_{1}^{2}c_{2}^{2}\left|a\right|^{2}\left|b\right|^{2}}.\]

Tracing out $E$-qubits in (\ref{Gl E1}) forces $\hat{\rho}_{SE_{L}}^{out}$
to acquire the form
\begin{equation}
\begin{array}{c}
\hat{\rho}_{SE}^{out}\left(L<n\right)=\left|a\right|^{2}\left|0\right\rangle \left\langle 0\right|\otimes\left|s_{c_{1}}^{L}\right\rangle \left\langle s_{c_{1}}^{L}\right|\\
+\left|b\right|^{2}\left|1\right\rangle \left\langle 1\right|\otimes\left|s_{c_{2}}^{L}\right\rangle \left\langle s_{c_{2}}^{L}\right|,
\end{array}\label{Gl E2}
\end{equation}
for which one in general has $H\left(S,\, E_{L<n}\right)>0$. Again,
without loss of generality, let us set in (\ref{Gl E2}) $L=n=1$:
w.r.t. (\ref{Gl E2}) $\hat{\rho}_{E}^{out}\left(L=n=1\right)$ remains
the same as in (\ref{Gl E1}), whereas $\hat{\rho}_{SE}^{out}\left(L=n=1\right)$
from (\ref{Gl E2}) leads $\forall\phi$ to non-zero eigenvalues \[ \lambda_{1}^{SE\left(L=n=1\right)}=\left|a\right|^{2}\,,\lambda_{2}^{SE\left(L=n=1\right)}=\left|b\right|^{2}.\]
Since $\left(c_{1},\, c_{2}\right)>0$ are parametrized by complementary
transcendent functions of the $\phi$-parameter, the only way to satisfy
the MI-plateau condition between $H\left(S,\, E_{L}\right)$ and $H\left(E_{L}\right)$
is to demand $H\left(S,\, E_{L=n=1}\right)=H\left(E_{L=n=1}\right)$,
which can be achieved only if we choose 
\begin{equation}
c_{1}^{2}=c_{2}^{2}=1/2,\label{Gl E3}
\end{equation}
which leads to $E$-eigenstates $\left\{ \left|s_{1}\right\rangle ,\,\left|s_{2}\right\rangle \right\} $
of the CNOT-transformation $\widehat{u}_{j}^{\left(\phi=\pi/2\right)}$
from (\ref{Gl 3.6}). Otherwise, $\forall\left(c_{1}\neq c_{2}\right)$
one has $H\left(S,\, E_{L=n=1}\right)>H\left(E_{L=n=1}\right)$. Thus,
(\ref{Gl E3}) shows that w.r.t. the $S$-pointer basis given by the
standard computational basis $\left\{ \left|\pi_{i}\right\rangle \right\} \equiv\left\{ \left|0\right\rangle ,\,\left|1\right\rangle \right\} $
solely the CNOT-transformation allows Quantum Darwinism to appear.

However, what happens if we generalize (\ref{Gl 3.6}) to an open $k>1$ qubit
system $S$? Since there are only two eigenstates $\left\{ \left|s_{1}\right\rangle ,\,\left|s_{2}\right\rangle \right\} $
of $\widehat{u}_{j}^{\left(\phi=\pi/2\right)}$ in (\ref{Gl 2.1})-(\ref{Gl 2.2}),
the easiest way to generalize (\ref{Gl E2})-(\ref{Gl E3}) to $k>1$
$S$-qubits is according to
\begin{align}
\label{Gl E4}
\begin{split}
\hat{\rho}_{SE}^{out}\left(L\right)=&\left(\underset{i=0}{\overset{2^{^{k-1}-1}}{\sum}}\left|a_{i}\right|^{2}\left|i\right\rangle \left\langle i\right|\right)\otimes\left|s_{1}^{L}\right\rangle \left\langle s_{1}^{L}\right|+\\
&\left(\underset{j=2^{^{k-1}}}{\overset{2^{^{k}}-1}{\sum}}\left|a_{j}\right|^{2}\left|j\right\rangle \left\langle j\right|\right)\otimes\left|s_{2}^{L}\right\rangle \left\langle s_{2}^{L}\right|,
\end{split}
\end{align}
w.r.t. an arbitrary probability distribution of an open system $S$ given by $1>\left|a_{i}\right|^{2}>0,\, i\in\left\{ 0,\,...,\,2^{k}-1\right\} $.
However, the eigenvalues of (\ref{Gl E4}), 
\begin{equation}
\begin{array}{c}
\begin{array}{c}
\hat{\rho}_{SE}^{out}\left(L\right):\,\left\{ \lambda_{i}^{SE}=\left|a_{i}\right|^{2}\right\} _{i=0}^{i=2^{k}-1}\Rightarrow\\
\left(1-\delta_{k,1}\right)\cdot H\left(E_{f}\right)\leq H\left(S,\, E_{f}\right)=H\left(S_{class}^{k\geq1}\right)\,\forall L
\end{array}\\
\begin{array}{c}
\hat{\rho}_{E}^{out}\left(L\right):\,\lambda_{1}^{E}=\underset{i=0}{\overset{2^{^{k-1}-1}}{\sum}}\left|a_{i}\right|^{2},\,\lambda_{2}^{E}=\underset{j=2^{^{k-1}}}{\overset{2^{^{k}}-1}{\sum}}\left|a_{j}\right|^{2}\\
\Rightarrow H\left(E_{f}\right)=-\underset{i=1}{\overset{2}{\sum}}\lambda_{i}^{E}\cdot\log_{2}\lambda_{i}^{E}\leq1\,\forall L,
\end{array}
\end{array}\label{Gl E5}
\end{equation}
indicate that QD appears for (\ref{Gl E4}) if and only if $k=1$,
yielding \[I\left(S:\, E_{f}\right)/H\left(S_{class}^{k\geq1}\right)=H\left(E_{f}\right)/
H\left(S_{class}^{k\geq1}\right)=1.\]

Accordingly, generalizing (\ref{Gl E1}) with (\ref{Gl E3}) as 
\begin{equation}
\begin{array}{c}
\left|\Psi_{SE}^{out}\left(L\right)\right\rangle =\underset{i=0}{\overset{2^{k-1}-1}{\sum}}a_{i}\left|i\right\rangle \left|s_{1}^{L}\right\rangle +\underset{j=2^{k-1}}{\overset{2^{k}-1}{\sum}}a_{j}\left|j\right\rangle \left|s_{2}^{L}\right\rangle \\
\underset{i=0}{\overset{2^{k}-1}{\sum}}\left|a_{i}\right|^{2}\overset{!}{=}1,\,\hat{\rho}_{SE}^{out}\left(L\right)=\left|\Psi_{SE}^{out}\left(L\right)\right\rangle \left\langle \Psi_{SE}^{out}\left(L\right)\right|,
\end{array}\label{Gl E6}
\end{equation}
where $H\left(E_{f}\right)=-\underset{i=1}{\overset{2}{\sum}}\lambda_{i}^{E}\log_{2}\lambda_{i}^{E}\leq1\,\forall L$
follows from (\ref{Gl E5}), $H\left(S\right)=H\left(S_{class}\right)$ and
$H\left(S,\, E_{f}\right)$ behaves in a two-fold way: 
\textbf{1)}
if $L=n\geq1$, (\ref{Gl E6}) is pure and we have the entropy relation $H\left(S,\, E_{f}\right)=0<H\left(E_{f}\right)$,
yielding \[I\left(S:\, E_{f}\right)=2H\left(S_{class}\right)\] (>>quantum
peak<<); 
\textbf{2)} For $1\leq L<n$ (\ref{Gl E6}) contains $2^{k-1}$
>>diagonal $S$-subspaces<<, half of which are organized according
to $\left|s_{1}^{L}\right\rangle \left\langle s_{1}^{L}\right|$,
whereas the remaining $2^{k-1}$ >>diagonal $S$-subspaces<< of
(\ref{Gl E6}) are ordered according to $\left|s_{2}^{L}\right\rangle \left\langle s_{2}^{L}\right|$.
This implies \[I\left(S:\, E_{f}\right)=H\left(S_{class}\right)\, \forall\left(1\leq L<n\right),\]
since $H\left(S,\, E_{f}\right)=H\left(E_{f}\right)=H\left(S_{class}^{k=1}\right)$,
as in (\ref{Gl E5}). 

In (\ref{Gl E6}) Quantum Darwinism does not
appear for $k>1$ and $1\leq L\leq n$, since $\hat{\rho}_{S}^{out}$
of (\ref{Gl E6}) has for $k\geq1$ only two eigenvalues \[\lambda_{1}^{S}=\underset{i=0}{\overset{2^{k-1}-1}{\sum}}\left|a_{i}\right|^{2},\,
\lambda_{2}^{S}=\underset{j=2^{k-1}}{\overset{2^{k}-1}{\sum}}\left|a_{j}\right|^{2}\]
(with $\lambda_{1}^{S}+\lambda_{2}^{S}\overset{!}{=}1$), corresponding
to eigenvalues of $k=1$ system $S$. Thus: if we organize $\hat{\rho}_{SE}^{out}\left(L\right)$
according to (\ref{Gl E6}), we could maximally store \[1\geq H\left(S_{class}\right)=-\lambda_{1}^{S}\log_{2}\lambda_{1}^{S}-
\lambda_{2}^{S}\log_{2}\lambda_{2}^{S}>0\]
of a $k=1$ system $S$, even if one should insist on $k>1$ (the
PIP for $k=1$ in (\ref{Gl E6}) is given by Fig. \ref{Fig.1}), i.e. in (\ref{Gl E6}) Quantum Darwinism appears only for $k=1$.

\section{Analytic reconstruction of attractor spaces\label{AAC}}

~~~In this section we intend to sketch how one can reconstruct
the maximal and minimal attractor spaces by utilizing
the QR-decomposition method.

\subsection{Maximal attractor space\label{AAC1}}

~~~The maximal attractor space and its basis states $\widehat{X}_{\lambda,i}$
of the random unitary evolution (\ref{Gl 2.1})-(\ref{Gl 2.5}) w.r.t.
a specific relevant eigenvalue $\lambda$ follow as a solution to
the eigenvalue equation (\ref{Gl 2.6}) obtained by means of the QR-decomposition
if we assume environment $E$ to contain mutually non-interacting
qubits. Since each directed edge of the ID in Fig. \ref{Fig.3} corresponds
to an additional linear equation (constraint) in (\ref{Gl 2.6}),
the minimal number of constraints (and thus the maximal attractor
space dimension $d_{n\geq k}^{\lambda}$) one could allow within the
random unitary evolution model is given by the so called Koenig-IDs
\cite{key-6}, in which only the $S$-qubits interact with $E$-qubits.
In the following we will first determine $d_{n\geq k}^{\lambda}$.

\subsubsection{Dimensionality\label{AAC1.1}}

~

By implementing the QR-decomposition numerically one notices for $n\geq k$
that within the maximal attractor space there are only two subspaces
with non-zero dimension $d^{\lambda}$ associated with eigenvalues
$\lambda=\pm1$ of (\ref{Gl 2.6}) \cite{key-1,key-2,key-4}. From
the numerically available data one can easily deduce for $n\geq k\geq1$
that the following dimension formulas hold: for the eigenvalue $\lambda=1$
\begin{equation}
\begin{array}{c}
d_{n}^{\lambda=1}=4^{n}+3\cdot2^{n}\left(2^{k}-1\right)+\\
\left(2^{k}-1\right)\left(2^{k}-2\right)+\delta_{n=k,1},
\end{array}\label{Gl G2}
\end{equation}
for the eigenvalue $\lambda=-1$
\begin{equation}
d_{n}^{\lambda=-1}=3\cdot2^{n}+3\cdot2^{k}-6-5\cdot\delta_{n=k,1}.\label{Gl G3}
\end{equation}
(\ref{Gl G2})-(\ref{Gl G3}) can be easily proven by induction. Furthermore,
one also sees from numerical data that for $k>n$ one has $d_{n<k}^{\lambda}=d_{n\leftrightarrow k}^{\lambda}\,\forall\lambda$,
i.e. $d_{n<k}^{\lambda}$ follows from $d_{n'=k\geq k'=n}^{\lambda}$
after interchanging $k$ with $n$ in (\ref{Gl G2})-(\ref{Gl G3}).

\subsubsection{State structure\label{AAC1.2}}

~

Implementing the QR-decomposition (s. \cite{key-5}) for IDs with
mutually non-interacting $E$-qubits and using the environmental $\widehat{u}_{j}^{\left(\phi\right)}$-symmetry
states $\left\{ \left|s_{c_{1}}^{L}\right\rangle ,\,\left|s_{c_{2}}^{L}\right\rangle \right\} $
from (\ref{Gl E1})-(\ref{Gl E2}) to classify the solutions (attractor
states) $\widehat{X}_{\lambda,i}$ of (\ref{Gl 2.6}) one obtains
$\forall\left(n\geq k\geq1\right)$ the following two attractor subspaces
associated with the two relevant eigenvalues $\lambda\in\left\{ 1,\,-1\right\} $:
\begin{equation}
\left.\begin{array}{c}
|0_{k}\rangle\langle x|\otimes|y\rangle\langle s_{c_{1}}^{n}|,\,|x\rangle\langle0_{k}|\otimes|s_{c_{1}}^{n}\rangle\langle y|\\
\begin{array}{c}
|0_{k}\rangle\langle0_{k}|\otimes|y\rangle\langle z|\\
\left|x\right\rangle \langle x|\underset{i=1}{\overset{n}{\otimes}}A_{\gamma_{i}},\,|x\rangle\langle w|\otimes|s_{c_{1}}^{n}\rangle\langle s_{c_{1}}^{n}|
\end{array}
\end{array}\right\} \lambda=1,\label{Gl G4}
\end{equation}
with \[
\begin{array}{c}
\left|\chi_{k}\right\rangle =\left|\chi\right\rangle ^{\otimes k},\,\left(\chi,\,\gamma_{i}\right)\in\left\{ 0,\,1\right\} \\
\ensuremath{\left(x,\, w\right)\in\left\{ 0,\,...,\,2^{k}-1\right\} },\,\ensuremath{\left(y,\, z\right)\in\left\{ 0,\,...,\,2^{n}-1\right\} }\\
\ensuremath{(x\neq w)\neq0_{k},\, A_{0}=|s_{c_{1}}\rangle\langle s_{c_{1}}|,\, A_{1}=\hat{I}-|s_{c_{1}}\rangle\langle s_{c_{1}}|}
\end{array}
\]
and
\begin{equation}
\left.\begin{array}{c}
|0_{k}\rangle\langle1_{k}|\otimes|y\rangle\langle s_{c_{2}}^{n}|,\,|1_{k}\rangle\langle0_{k}|\otimes|s_{c_{2}}^{n}\rangle\langle y|\\
|1_{k}\rangle\langle1_{k}|\underset{i=1}{\overset{n}{\otimes}}B_{\gamma_{i}},\,|x\rangle\langle x|\otimes|s_{c_{2}}^{n}\rangle\langle s_{c_{2}}^{n}|\\
|x\rangle\langle1_{k}|\otimes|s_{c_{1}}^{n}\rangle\langle s_{c_{2}}^{n}|,\,|1_{k}\rangle\langle x|\otimes|s_{c_{2}}^{n}\rangle\langle s_{c_{1}}^{n}|
\end{array}\right\} \lambda=-1,\label{Gl G5}
\end{equation}
with \[
\begin{array}{c}
x\neq(0_{k},\,1_{k}),\, B_{0}=\left(\sqrt{2}\right)^{-1}\left(|0\rangle\langle1|-|1\rangle\langle0|\right)\\
\ensuremath{B_{1}=\left(\sqrt{2}\right)^{-1}\left[-\sin\phi|0\rangle\langle0|+\sin\phi|1\rangle\langle1|+\cos\phi|0\rangle\langle1|+\right.}\\
\ensuremath{\left.\cos\phi|1\rangle\langle0|\right]}.
\end{array}
\]
(\ref{Gl G4})-(\ref{Gl G5}) are in accord with (\ref{Gl G2})-(\ref{Gl G3})
and contain orthonormalized attractor states $\widehat{X}_{\lambda,i}$,
with \[\left\langle \widehat{X}_{\lambda,i},\,\widehat{X}_{\lambda',i'}\right\rangle :=\delta_{\lambda,\,\lambda'}\delta_{i,\, i'},\]
given by the Hilbert-Schmidt scalar product \[\left\langle \widehat{X}_{\lambda,i},\,\widehat{X}_{\lambda',i'}\right\rangle :=\textrm{Tr}\left[\widehat{X}_{\lambda,i}\left(\widehat{X}_{\lambda',i'}
\right)^{\dagger}\right].\]

\subsection{Minimal attractor space\label{AAC2}}

~~~Now we turn our attention to environments $E$ whose all $n$
qubits are allowed to mutually interact with each other, as depicted
by the ID in Fig. \ref{Fig.3} and already studied in \cite{key-1,key-2,key-4}.

\subsubsection{Dimensionality\label{AAC2.1}}

~

From \cite{key-1,key-2,key-4} we know that $E$ enclosing mutually
via $\widehat{u}_{j}^{\left(\phi\right)}$ interacting $n$ qubits
(with $n\geq k\geq1$) leads to the the most constrained (strongly
connected) ID with an attractor subspace associated with the eigenvalue
$\lambda=1$ of (\ref{Gl 2.6}) of minimal dimension
\begin{equation}
d_{n}^{\lambda=1,min}=4^{k}+3\cdot2^{k}+1,\label{Gl G6}
\end{equation}
whereas the dimensionality of the $\lambda=-1$ attractor subspace
satisfies
\begin{equation}
d_{n}^{\lambda=-1,min}=\left\{ \begin{array}{c}
1\,\textrm{if}\, n=k=1\\
0,\,\textrm{otherwise}.
\end{array}\right.\label{Gl G7}
\end{equation}
Since Quantum Darwinism involves environments $E$ with $n\gg1$ qubits,
we may conclude that within the minimal attractor space only the $\lambda=1$
subspace contributes to the evolution of $\hat{\rho}_{SE}^{in}$.

\subsubsection{State structure\label{AAC2.2}}

~

From \cite{key-1,key-2,key-4} we know that (\ref{Gl G6}) corresponds
to the following structure of the linear independent (however not
yet orthonormalized) $\widehat{X}_{\lambda=1,i}$-states

\begin{equation}
\begin{array}{c}
\left|x\right\rangle \left\langle x\right|\otimes\hat{\textrm{I}}_{n},\,\left|0_{k}\right\rangle \left\langle x\right|\otimes\left|0_{n}\right\rangle \left\langle s_{c_{1}}^{n}\right|,\,\left|x\right\rangle \left\langle 0_{k}\right|\otimes\left|s_{c_{1}}^{n}\right\rangle \left\langle 0_{n}\right|\\
\left|x\right\rangle \left\langle y\right|\otimes\left|s_{c_{1}}^{n}\right\rangle \left\langle s_{c_{1}}^{n}\right|,\,\left|0_{k}\right\rangle \left\langle 0_{k}\right|\otimes\left|0_{n}\right\rangle \left\langle 0_{n}\right|,
\end{array}\label{Gl G8}
\end{equation}
whereas (\ref{Gl G7}) corresponds for $k=n=1$ to the only non-zero
orthonormalized $\widehat{X}_{\lambda=-1,i}$-state 
\begin{equation}
\begin{array}{c}
\hat{X}_{\lambda=-1,\, i=1}^{n=k=1}=\frac{1}{\sqrt{6}}\left(\left|01\right\rangle \left\langle 11\right|-\left|10\right\rangle \left\langle 11\right|\right.\\
\left.-\left|11\right\rangle \left\langle 01\right|+\left|11\right\rangle \left\langle 10\right|-\left|01\right\rangle \left\langle 10\right|+\left|10\right\rangle \left\langle 01\right|\right),
\end{array}\label{Gl G9}
\end{equation}
with 

\begin{align*}
\begin{split}
\left(x,\, y\right)&\in\left\{ 0,\,...,\,2^{k}-1\right\} \\
\textrm{Tr}_{E}\left[\hat{\textrm{I}}_{n}\right]&=2^{n}\\ \hat{\textrm{I}}_{n}&=\hat{\textrm{I}}_{1}^{\otimes n}\\
\end{split}
\end{align*}
and $\left\{ \left|s_{c_{1}}^{L}\right\rangle ,\,\left|s_{c_{2}}^{L}\right\rangle \right\} $
from (\ref{Gl E1})-(\ref{Gl E2}). However, (\ref{Gl G9}) does not
contribute to the evolution of $\hat{\rho}_{SE}^{in}$ from the point
of view of Quantum Darwinism, which necessitates us to start with
$\hat{\rho}_{SE}^{in}$ enclosing environments $E$ with $n\gg k\geq1$
qubits.

\newpage
\twocolumn[{%

\section{Output states $\hat{\rho}_{SE_{L}}^{out}$ of the random unitary
evolution used in section \ref{A3} \label{AAD_ges}}

~~~In this appendix we list the output states $\hat{\rho}_{SE_{L}}^{out}$
of the random unitary evolution used in section \ref{A3}.

\subsection{Output states $\hat{\rho}_{SE_{L}}^{out}$ of the random unitary
evolution for the maximal attractor space\label{AAD1-1}}

~~~In this appendix we list the output states $\hat{\rho}_{SE_{L}}^{out}$
of the random unitary evolution used in section \ref{A3} of the main
text when discussing Quantum Darwinism from the point of view of the
maximal attractor space.

\paragraph{I) Input: $\hat{\rho}_{SE}^{in}=\left|\Psi_{S}^{in}\right\rangle \left\langle \Psi_{S}^{in}\right|\otimes\hat{\rho}_{E}^{in}$,
$\left|\Psi_{S}^{in}\right\rangle =\protect\underset{m=0}{\protect\overset{2^{k}-1}{\sum}}a_{m}\left|m\right\rangle $,
$\hat{\rho}_{E}^{in}=\left|z^{L=n}\right\rangle \left\langle z^{L=n}\right|$,
$\widehat{u}_{j}^{\left(\phi=\pi/2\right)}$-evolution}

~

\begin{equation}
\begin{array}{c}
\hat{\rho}_{SE}^{out}=\left|\Psi'\right\rangle \left\langle \Psi'\right|+2^{-n}\underset{m=1}{\overset{2^{k}-1}{\sum}}\left|a_{m}\right|^{2}\left|m\right\rangle \left\langle m\right|\otimes\Biggl(\widehat{I}_{n}-\left|s_{1}^{n}\right\rangle \left\langle s_{1}^{n}\right|\Biggr)\\
+\left(-1\right)^{N}\cdot\left\{ 2^{-n/2}\left(-1\right)^{M}\Biggl[a_{0}a_{2^{k}-1}^{*}\left|0\right\rangle \left\langle 2^{k}-1\right|\otimes\left|z^{n}\right\rangle \left\langle s_{2}^{n}\right|+a_{2^{k}-1}a_{0}^{*}\left|2^{k}-1\right\rangle \left\langle 0\right|\otimes\left|s_{2}^{n}\right\rangle \left\langle z^{n}\right|\Biggr]\right.\\
+2^{-n}\underset{m=1}{\overset{2^{k}-2}{\sum}}\left|a_{m}\right|^{2}\left|m\right\rangle \left\langle m\right|\otimes\left|s_{2}^{n}\right\rangle \left\langle s_{2}^{n}\right|+\left(-1\right)^{n}\cdot2^{-n/2}\left|a_{2^{k}-1}\right|^{2}\left|2^{k}-1\right\rangle \left\langle 2^{k}-1\right|\underset{i=1}{\overset{n}{\otimes}}\hat{B}_{1}^{\pi/2}\\
\left.+2^{-n}\underset{m=1}{\overset{2^{k}-2}{\sum}}a_{m}a_{2^{k}-1}^{*}\left|m\right\rangle \left\langle 2^{k}-1\right|\otimes\left|s_{1}^{n}\right\rangle \left\langle s_{2}^{n}\right|+2^{-n}\underset{m=1}{\overset{2^{k}-2}{\sum}}a_{2^{k}-1}a_{m}^{*}\left|2^{k}-1\right\rangle \left\langle m\right|\otimes\left|s_{2}^{n}\right\rangle \left\langle s_{1}^{n}\right|\right\} 
\end{array},\label{Gl I1}
\end{equation}
where $z^{L=n}\in\left\{ 0,\,...,\,2^{n}-1\right\} $, $\left\{ \left|s_{1}\right\rangle ,\,\left|s_{2}\right\rangle \right\} $
as in (\ref{Gl 3.6}), $\left\langle s_{1}\right|\xi\rangle=2^{-1/2}$,
$\left\langle s_{2}\right|\xi\rangle=\left(-1\right)^{\xi}2^{-1/2}$
for $\xi\in\left\{ 0,\,1\right\} $, $\left|\Psi'\right\rangle =a_{0}\left|0\right\rangle \otimes\left|z^{n}\right\rangle +\underset{m=1}{\overset{2^{k}-1}{\sum}}a_{m}2^{-n/2}\left|m\right\rangle \otimes\left|s_{1}^{n}\right\rangle $,
$\hat{B}_{1}^{\pi/2}=\left(\sqrt{2}\right)^{-1}\left[\left|1\right\rangle \left\langle 1\right|-\left|0\right\rangle \left\langle 0\right|\right]$,
and $M$ is the number of $\left|1\right\rangle $-one qubit states
in $\left|z^{L=n}\right\rangle $.

\paragraph{II) Input: $\hat{\rho}_{SE}^{in}=\left|\Psi_{S}^{in}\right\rangle \left\langle \Psi_{S}^{in}\right|\otimes\hat{\rho}_{E}^{in}$,
$\left|\Psi_{S}^{in}\right\rangle =\protect\underset{m=0}{\protect\overset{2^{k=1}-1}{\sum}}a_{m}\left|m\right\rangle $,
$\hat{\rho}_{E}^{in}=2^{-1}\left(\left|0_{n}\right\rangle \left\langle 0_{n}\right|+\left|1_{n}\right\rangle \left\langle 1_{n}\right|\right)$,
$\widehat{u}_{j}^{\left(\phi=\pi/2\right)}$-evolution}

\begin{equation}
\begin{array}{c}
\hat{\rho}_{SE}^{out}=\left|a_{0}\right|^{2}\left|0\right\rangle \left\langle 0\right|\otimes\hat{\rho}_{E}^{in}+\left|a_{1}\right|^{2}\left|1\right\rangle \left\langle 1\right|\otimes2^{-n}\hat{I}_{n}+\\
2^{-n/2-1}\cdot\left(a_{0}a_{1}^{*}\left|0\right\rangle \left\langle 1\right|\otimes\Biggl[\left|0_{n}\right\rangle \left\langle s_{1}^{n}\right|+\left|1_{n}\right\rangle \left\langle s_{1}^{n}\right|\Biggr]+a_{1}a_{0}^{*}\left|1\right\rangle \left\langle 0\right|\otimes\Biggl[\left|s_{1}^{n}\right\rangle \left\langle 0_{n}\right|+\left|s_{1}^{n}\right\rangle \left\langle 1_{n}\right|\Biggr]\right)\\
+\left(-1\right)^{N}\Biggl\{2^{-n/2-1}\Biggl(a_{0}a_{1}^{*}\left|0\right\rangle \left\langle 1\right|\otimes\Biggl[\left|0_{n}\right\rangle \left\langle s_{2}^{n}\right|+\left(-1\right)^{n}\left|1_{n}\right\rangle \left\langle s_{2}^{n}\right|\Biggr]+\\
a_{1}a_{0}^{*}\left|1\right\rangle \left\langle 0\right|\otimes\Biggl[\left|s_{2}^{n}\right\rangle \left\langle 0_{n}\right|+\left(-1\right)^{n}\left|s_{2}^{n}\right\rangle \left\langle 1_{n}\right|\Biggr]\Biggr)+\left.2^{-1}\left|a_{1}\right|^{2}\left|1\right\rangle \left\langle 1\right|2^{-n/2}\left(1+\left(-1\right)^{n}\right)\underset{i=1}{\overset{n}{\otimes}}\hat{B}_{1}^{\pi/2}\right\} 
\end{array}\label{Gl I3}
\end{equation}

\paragraph{III) Input: $\hat{\rho}_{SE}^{in}=\left|\Psi_{S}^{in}\right\rangle \left\langle \Psi_{S}^{in}\right|\otimes\hat{\rho}_{E}^{in}$,
$\left|\Psi_{S}^{in}\right\rangle =\protect\underset{m=0}{\protect\overset{2^{k=1}-1}{\sum}}a_{m}\left|m\right\rangle $,
$\hat{\rho}_{E}^{in}=2^{-n}\hat{I}_{n}$, $\widehat{u}_{j}^{\left(\phi=\pi/2\right)}$-evolution}

\begin{equation}
\begin{array}{c}
\hat{\rho}_{SE}^{out}\left(L=n\right)=\left(\left|a_{0}\right|^{2}\left|0\right\rangle \left\langle 0\right|+\left|a_{1}\right|^{2}\left|1\right\rangle \left\langle 1\right|\right)\otimes2^{-n}\hat{I}_{n}+\\
\left(a_{0}a_{1}^{*}\left|0\right\rangle \left\langle 1\right|+a_{1}a_{0}^{*}\left|1\right\rangle \left\langle 0\right|\right)\otimes2^{-n}\left(\left|s_{1}^{L=n}\right\rangle \left\langle s_{1}^{L=n}\right|+\left(-1\right)^{N}\left|s_{2}^{L=n}\right\rangle \left\langle s_{2}^{L=n}\right|\right)
\end{array}\label{Gl I4}
\end{equation}

\paragraph{IV) Input: $\hat{\rho}_{SE}^{in}=\left|\Psi_{S}^{in}\right\rangle \left\langle \Psi_{S}^{in}\right|\otimes\hat{\rho}_{E}^{in}$,
$\left|\Psi_{S}^{in}\right\rangle =\protect\underset{m=0}{\protect\overset{2^{k=1}-1}{\sum}}a_{m}\left|m\right\rangle $,
$\hat{\rho}_{E}^{in}=2^{-1}\left(\left|0_{n}\right\rangle \left\langle 0_{n}\right|+\left|10_{n-1}\right\rangle \left\langle 10_{n-1}\right|\right)$,
$\widehat{u}_{j}^{\left(\phi=\pi/2\right)}$-evolution}

~

$\hat{\rho}_{SE}^{out}\left(L=n\right)$ emerges from (\ref{Gl I3})
by applying the following substitutions:

\begin{equation}
\begin{array}{c}
\left|1_{n}\right\rangle \leftrightarrow\left|10_{n-1}\right\rangle,\,
\left(-1\right)^{n}\left(\left|1_{n}\right\rangle \left\langle s_{2}^{L}\right|+h.c.\right)\leftrightarrow\left(-1\right)^{1}\left(\left|10_{n-1}\right\rangle \left\langle s_{2}^{L}\right|+h.c\right)\\
\left(1+\left(-1\right)^{n}\right)\underset{i=1}{\overset{n}{\otimes}}\hat{B}_{1}^{\pi/2}\cdot\delta_{L,n}\leftrightarrow\left(-1\right)^{n-1}\underset{i=1}{\overset{n}{\otimes}}\hat{B}_{1}^{\pi/2}\cdot\delta_{L,n}
\end{array}\label{Gl I5}
\end{equation}
}] %End of twocolumn
\newpage
\twocolumn[{%

\paragraph{V) Input: $\left|\Psi_{SE}^{in}\right\rangle =a\left|0_{k=1}\right\rangle \otimes\left|s_{1}^{L=n}\right\rangle +b\left|1_{k=1}\right\rangle \otimes\left|s_{2}^{L=n}\right\rangle $,
$\widehat{u}_{j}^{\left(\phi=\pi/2\right)}$-evolution}

\begin{equation}
\left|\Psi_{SE_{L=n}}^{out}\right\rangle =a\left|0\right\rangle \otimes\left|s_{1}^{n}\right\rangle +\left(-1\right)^{N}b\left|1\right\rangle \otimes\left|s_{2}^{n}\right\rangle \label{Gl I6}
\end{equation}

\subsection{Output states $\hat{\rho}_{SE}^{out}$ of the random unitary evolution
for the minimal attractor space\label{AAD2-1}}

~~~In this appendix we list the output states $\hat{\rho}_{SE_{L}}^{out}$
of the random unitary evolution used in section \ref{A3} of the main
text when discussing Quantum Darwinism from the point of view of the
minimal $\lambda=1$ attractor subspace.

\paragraph{I) Input: $\hat{\rho}_{SE}^{in}=\left|\Psi_{S}^{in}\right\rangle \left\langle \Psi_{S}^{in}\right|\otimes\hat{\rho}_{E}^{in}$,
$\left|\Psi_{S}^{in}\right\rangle =\protect\underset{m=0}{\protect\overset{2^{k}-1}{\sum}}a_{m}\left|m\right\rangle $,
$\hat{\rho}_{E}^{in}=\left|0_{n}\right\rangle \left\langle 0_{n}\right|$,
$\widehat{u}_{j}^{\left(\phi=\pi/2\right)}$-evolution}

\begin{equation}
\widehat{\rho}_{SE}^{out}=\left|\Psi'\right\rangle \left\langle \Psi'\right|+2^{-n}\cdot\underset{y=1}{\overset{2^{k}-1}{\sum}}\left|a_{y}\right|^{2}\left|y\right\rangle \left\langle y\right|\otimes\Biggl(\widehat{I}_{n}-\left|s_{1}^{n}\right\rangle \left\langle s_{1}^{n}\right|\Biggr),\label{Gl J1}
\end{equation}
where $\left|\Psi'\right\rangle =a_{0}\left|0\right\rangle \otimes\left|0_{n}\right\rangle +\underset{y=1}{\overset{2^{k}-1}{\sum}}a_{y}2^{-n/2}\left|y\right\rangle \otimes\left|s_{1}^{n}\right\rangle $.

\paragraph{II) Input: $\hat{\rho}_{SE}^{in}=\left|\Psi_{S}^{in}\right\rangle \left\langle \Psi_{S}^{in}\right|\otimes\hat{\rho}_{E}^{in}$,
$\left|\Psi_{S}^{in}\right\rangle =\protect\underset{m=0}{\protect\overset{2^{k}-1}{\sum}}a_{m}\left|m\right\rangle $,
$\hat{\rho}_{E}^{in}=\left|1_{n}\right\rangle \left\langle 1_{n}\right|$,
$\widehat{u}_{j}^{\left(\phi=\pi/2\right)}$-evolution}

~

\begin{equation}
\begin{array}{c}
\widehat{\rho}_{SE}^{out}=\left|a_{0}\right|^{2}\cdot2^{-n}\left(1-2^{-n}\right)^{-1}\left|0\right\rangle \left\langle 0\right|\otimes\left(\widehat{I}_{n}\cdot2^{-n}-\left|0_{n}\right\rangle \left\langle 0_{n}\right|\right)+2^{-n}\cdot\underset{y=0}{\overset{2^{k}-1}{\sum}}\left|a_{y}\right|^{2}\left|y\right\rangle \left\langle y\right|\otimes\widehat{I}_{n}\\
-2^{-3n/2}\cdot\left(1-2^{-n}\right)^{-1}\underset{y=1}{\overset{2^{k}-1}{\sum}}\left\{ a_{0}a_{y}^{*}\left|0\right\rangle \left\langle y\right|\otimes\left|0_{n}\right\rangle \left\langle s_{1}^{n}\right|+a_{y}a_{0}^{*}\left|y\right\rangle \left\langle 0\right|\otimes\left|s_{1}^{n}\right\rangle \left\langle 0_{n}\right|\right\} \\
+2^{-n}\cdot\underset{\left(x\neq y\right)=1}{\overset{2^{k}-1}{\sum}}\left\{ \left(1-2^{-n}\right)^{-1}\cdot a_{0}a_{y}^{*}\left|0\right\rangle \left\langle y\right|+\left(1-2^{-n}\right)^{-1}\cdot a_{y}a_{0}^{*}\left|y\right\rangle \left\langle 0\right|+a_{y}a_{x}^{*}\left|y\right\rangle \left\langle x\right|\right\} \otimes\left|s_{1}^{n}\right\rangle \left\langle s_{1}^{n}\right|
\end{array}\label{Gl J2}
\end{equation}

\paragraph{III) Input: $\hat{\rho}_{SE}^{in}=\left|\Psi_{S}^{in}\right\rangle \left\langle \Psi_{S}^{in}\right|\otimes\hat{\rho}_{E}^{in}$,
$\left|\Psi_{S}^{in}\right\rangle =\protect\underset{m=0}{\protect\overset{2^{k=1}-1}{\sum}}a_{m}\left|m\right\rangle $,
$\hat{\rho}_{E}^{in}=2^{-1}\left(\left|0_{n}\right\rangle \left\langle 0_{n}\right|+\left|1_{n}\right\rangle \left\langle 1_{n}\right|\right)$,
$\widehat{u}_{j}^{\left(\phi=\pi/2\right)}$-evolution}

~

\begin{equation}
\begin{array}{c}
\widehat{\rho}_{SE}^{out}=\left|a_{0}\right|^{2}\left|0\right\rangle \left\langle 0\right|\left(1-2^{-n}\right)^{-1}\otimes\left\{ \widehat{I}_{n}\cdot2^{-n-1}+\left|0_{n}\right\rangle \left\langle 0_{n}\right|\left(1-2^{1-n}\right)^{-1}\left(\frac{1}{2}-2^{1-n}+2^{1-2n}\right)\right\} \\
+\left|a_{1}\right|^{2}\left|1\right\rangle \left\langle 1\right|\otimes2^{-n}\cdot\widehat{I}_{n}+a_{0}a_{1}^{*}2^{-1}\left(1-2^{1-n}\right)^{-1}\left|0\right\rangle \left\langle 1\right|\otimes\left\{ \left|0_{n}\right\rangle \left\langle s_{1}^{n}\right|\cdot2^{-n/2}\left(1-2^{1-n}\right)+2^{-n}\left|s_{1}^{n}\right\rangle \left\langle s_{1}^{n}\right|\right\} \\
+a_{1}a_{0}^{*}2^{-1}\left(1-2^{1-n}\right)^{-1}\left|1\right\rangle \left\langle 0\right|\otimes\left\{ \left|s_{1}^{n}\right\rangle \left\langle 0_{n}\right|\cdot2^{-n/2}\left(1-2^{1-n}\right)+2^{-n}\left|s_{1}^{n}\right\rangle \left\langle s_{1}^{n}\right|\right\} 
\end{array}\label{Gl J3}
\end{equation}

\paragraph{IV) Input: $\hat{\rho}_{SE}^{in}=\left|\Psi_{S}^{in}\right\rangle \left\langle \Psi_{S}^{in}\right|\otimes\hat{\rho}_{E}^{in}$,
$\left|\Psi_{S}^{in}\right\rangle =\protect\underset{m=0}{\protect\overset{2^{k}-1}{\sum}}a_{m}\left|m\right\rangle $,
$\hat{\rho}_{E}^{in}=2^{-n}\hat{I}_{n}$, $\widehat{u}_{j}^{\left(\phi=\pi/2\right)}$-evolution}

~

\begin{equation}
\widehat{\rho}_{SE}^{out}=\underset{x=0}{\overset{2^{k}-1}{\sum}}\left|a_{x}\right|^{2}\left|x\right\rangle \left\langle x\right|\otimes\hat{\rho}_{E}^{in}+2^{-n}\underset{\left(x\neq y\right)=0}{\overset{2^{k}-1}{\sum}}a_{x}a_{y}^{*}\left|x\right\rangle \left\langle y\right|\otimes\left|s_{1}^{n}\right\rangle \left\langle s_{1}^{n}\right|\label{Gl J4}
\end{equation}

\paragraph{V) Input: $\left|\Psi_{SE}^{in}\right\rangle =a\left|0_{k=1}\right\rangle \otimes\left|s_{1}^{L}\right\rangle +b\left|1_{k=1}\right\rangle \otimes\left|s_{2}^{L}\right\rangle $,
$\widehat{u}_{j}^{\left(\phi=\pi/2\right)}$-evolution}

~

\begin{equation}
\widehat{\rho}_{SE}^{out}=\left|a_{0}\right|^{2}\left|0\right\rangle \left\langle 0\right|\otimes\left|s_{1}^{n}\right\rangle \left\langle s_{1}^{n}\right|+\left|a_{1}\right|^{2}\left(2^{n}-1\right)^{-1}\left|1\right\rangle \left\langle 1\right|\otimes\left(\widehat{I}_{n}-\left|s_{1}^{n}\right\rangle \left\langle s_{1}^{n}\right|\right),\label{Gl J5}
\end{equation}
}] %End of twocolumn

%--------------------REFERENCES--------------------

\end{document}